\documentclass[12pt, fleqn]{article}
\usepackage[utf8]{inputenc}
\usepackage{fancyhdr}
\usepackage[a4paper, total={6.5in, 9in}]{geometry}

\fancypagestyle{plain}{%
	\fancyhf{}%
}
\usepackage[onehalfspacing]{setspace}

\usepackage{authblk}
\title{A Network-Based Explanation of Inequality Perceptions}
\author[1]{Jan Schulz} 
\affil[1] {Economics Department, University of Bamberg}
\author[2]{Daniel M. Mayerhoffer}
\affil[2]{Institute for Political Science, University of Bamberg, Funded by the Deutsche Forschungsgemeinschaft (DFG, German Research Foundation) - 430621735}
\author[3]{Anna Gebhard}
\affil[3]{MathOpt Research Group, University of Magdeburg}

\usepackage{mathtools}
\usepackage{tikz}

\usepackage{multirow}
\usepackage{slashbox}
\usepackage{graphicx}

\usepackage{endnotes}
\makeatletter
\renewcommand*\makeenmark{\hbox{\textsuperscript{\@Alph{\theenmark}}}}
\makeatother

\usepackage{float}

\usepackage{amssymb}
\usepackage{amsmath}

\DeclareMathOperator*{\argmin}{arg\,min}

\usepackage{pgfplots}
\pgfplotsset{compat=1.6}
\usepackage{graphicx}
\usepackage{caption}
\usepackage{subcaption}
\usepackage{grffile}

\usepackage{graphicx} %Loading the package

\usepackage{chngcntr}
\counterwithout{figure}{section}
\usepackage{natbib}
\setcitestyle{authoryear,round,aysep={}}

\newenvironment{HD}[2][Heuristic Derivation]{\begin{trivlist}
		\item[\hskip \labelsep {\bfseries #1.}\hskip \labelsep {\bfseries}]}{\end{trivlist}}

\newenvironment{PS}[2][Proof Sketch]{\begin{trivlist}
		\item[\hskip \labelsep {\bfseries #1.}\hskip \labelsep {\bfseries}]}{\end{trivlist}}

\begin{document}
\maketitle 
\renewcommand*{\thefootnote}{\fnsymbol{footnote}}
\footnotetext[1]{Daniel Mayerhoffer's contribution was funded by the Deutsche Forschungsgemeinschaft (DFG, German Research Foundation) - 430621735. Furthermore, financial support for the paper by the University of Bamberg through the Fres(c)h grant no. 06999902 is gratefully acknowledged. We would also like to express our gratitude to two anonymous reviewers and Martin Everett for his excellent editorial work, Arndt Leininger and his students, Miriam B\"omer, Bettina Gregg, Johannes Marx, Moritz Schulz, Eleonora Priori and Jan Weber as well as the participants of the Networks 2021 conference, the 9th PhD conference for the renewal of constitutional economics, the 9th ECINEQ Meeting, the 33rd annual EAEPE conference, the annual PhD conference by the Hans B\"ockler Foundation 2021, the 2nd Scientific Workshop by the Network for Pluralist Economics, the 8th Workshop on Complexity, Innovation and Knowledge (WICK), the Colloquium for the Advancement of Knowledge in Economics (CAKE) of the University of Utah and 28th DVPW congress, especially our designated discussants Macartan Humphreys, Claudius Gr\"abner and Pietro Terna which were of great help at crucial junctions of this investigation. Finally, we also thank Carsten K\"allner for his able research assistance. All remaining errors are, of course, ours.}
\renewcommand*{\thefootnote}{\arabic{footnote}}

%%%%%%%%%%%%%%%%%%%%%%%%%%%%%%%%%%%%%%%%%%%%%%

% Abstract and keywords

\begin{abstract}
%and many other macroeconomic variables such as inflation or unemployment rates	
Across income groups and countries, individual citizens perceive economic inequality spectacularly wrong. These misperceptions have far-reaching consequences, as it might be \emph{perceived} inequality, not \emph{actual} inequality informing redistributive preferences. The prevalence of this phenomenon is independent of social class and welfare regime, which suggests the existence of a common mechanism behind public perceptions. The literature has identified several stylised facts on how individual perceptions respond to actual inequality and how these biases vary systematically along the income distribution. We propose a network-based explanation of perceived inequality building on recent advances in random geometric graph theory. The generating mechanism can replicate all of aforementioned stylised facts simultaneously. It also produces social networks that exhibit salient features of real-world networks; namely, they cannot be statistically distinguished from small-world networks, testifying to the robustness of our approach. Our results, therefore, suggest that homophilic segregation is a promising candidate to explain inequality perceptions with strong implications for theories of consumption and voting behaviour.

\end{abstract}

%%%%%%%%%%%%%%%%%%%%%%%%%%%%%%%%%%%%%%%%%%%%%%
% Start of  paragraph numbering. Please leave this untouched
%\parano{}

%%%%%%%%%%%%%%%%%%%%%%%%%%%%%%%%%%%%%%%%%%%%%%

\section{Introduction}\label{sec:Introduction}
Conventional modern macroeconomics has long recognised the crucial relevance of expectations and belief-formation for aggregate dynamics \citep{gali2015}. In particular, beliefs about economic inequality and perceptions of social hierarchy can inform individuals in such diverse fields as consumption decisions \citep{Duesenberry1949, veblen2001, Frank2014}, redistributive preferences and voting behaviour \citep{Gimpelson2018, Kim2018, Choi2019} or subjective well-being and ethical convictions \citep{kuhn2019, clark2010}. Even in the most sophisticated behavioural models, belief-formation is, however, typically either assumed to be atomistic \citep{gabaix2020} or does not systematically account for the impact of individual embeddedness within heterogeneous social contexts on those beliefs, even if social interaction is explicitly modelled \citep{flieth2002,lux2009}. We propose a parsimonious network-based model for the interaction of macro-level inequality, micro-level beliefs and the mediating effects of heterogeneous social contexts. In contrast to the assumption of deductive reasoning in orthodox models, we build on the empirically well-established notion that economic agents reason \emph{inductively} and generalise from finite samples. Recent theoretical and empirical work has demonstrated the potency of this approach in explaining phenomena in such diverse fields as human probability assessment \citep{sanborn2016, chater2020} or regional inequality \citep{collier2020}. The model is both consistent with several stylised facts about inequality perceptions and the micro-level evidence on the composition of social networks. 

The relevance of individual beliefs is perhaps best exemplified by spelling out its political economy implications. Across income groups and  countries, the public perception of economic inequality and many other macroeconomic variables is empirically wrong, often spectacularly so. Errors in those beliefs might be due to conceptually different problems: \emph{un}informed beliefs or \emph{mis}informed beliefs \citep{kuklinski2000}. Uninformed voters are ignorant about the actual state of affairs, while misinformed voters' beliefs are consistently deviating from it in one direction. The distinction is a crucial one. Uninformed voters' beliefs would cluster around the actual state of affairs and, with no systematic deviations, be correct in expectations. For uninformed voters, we only need one informed voter to tip elections under majority rule into the correct direction; a majority of ignorant individuals might nevertheless vote for the correct policy, which is now known as the 'miracle of aggregation' \citep{page1993}. However, his miraculous aggregation breaks down when we consider misinformed rather than uninformed voters with beliefs that are no longer randomly distributed but consistently tend in a (false) direction \citep{caplan2011}. The type of error in perceptions is thus intimately linked to the efficacy of democratic systems. For inequality perceptions, beliefs appear to be indeed the result of misinformation in this technical sense and they are consistently biased across income groups and welfare regimes. 

In contrast to much of the behavioural literature, we refrain from ad-hoc assumptions about possible biases, e.g., assuming that individuals tend to perceive themselves in the middle of social hierarchies \citep[cf., e.g.][]{Knell2020}. Instead, we assume unbiased information processing capabilities for all the economic agents. Information is, however, asymmetric and agents form estimates about aggregate variables according to their local information. We show that a parsimonious process can generate sufficiently skewed information sets to replicate the aforementioned stylised facts and generate perceived inequality levels that are quantitatively in line with recent empirical evidence for a large sample of $32$ OECD countries \citep{Choi2019}. In essence, we assume that agents (correctly) observe inequality within their local social network and (correctly) form estimates about the total population from them but still generate biased perceptions due to their network contacts not being representative for the overall population. Employing a new variant of a random geometric graph network, the assumption of income homophily alone can generate substantial misperception in line with the empirical evidence. The derived network topology also corresponds to empirically observed social networks across the world and features a small-world structure. Given the ubiquity of these topological features, our homophilic process appears to be a plausible candidate to explain the equally ubiquituous inequality misperceptions.

Our contribution is thus threefold: Firstly, we compile a list of four stylised empirical facts about income inequality and its biased perception from the nascent literature on the matter. Secondly, we develop a model that simultaneously replicates these stylised facts building on homophilic linkage and unbiased individual estimates based on local signals. Our model is quantitatively consistent with empirical estimates regarding both the input income distribution and the output perceptions, in contrast to the somewhat stylised models in the extant literature that also fail to replicate the dynamic behaviour of perceptions in response to changes in actual inequality. Thirdly, the network-formation algorithm presents a novel way of generating random geometric graph types of networks which is more intuitive for many application scenarios and allows specifying a minimum degree.

The remainder of this paper is organised as follows: Section 2 extracts four stylised empirical facts about inequality perception; furthermore, it reviews the evidence on empirical network topologies and individual belief formation within networks. Section 3 introduces the basic model of homophilic graph formation, reviews the main mechanisms generating heterogeneity in information sets, abd validates the model. Section 4 presents our analytic and simulative results, shows that they are consistent with the outlined stylised facts regarding network topologies as well as inequality perceptions and derives some important implications regarding heterogeneous segregation patterns across the income distribution. Section 5 concludes and discusses several promising avenues for further research, especially regarding consumption dynamics and voting behaviour.

\section{Related Literature}\label{sec:Related Literature}
%NEU!
Our model joins three different strands of literature. Empirical findings on inequality perceptions that a single theory or model has not yet explained constitute its main explanandum. As explanans, we develop a network model featuring the current state of research into both the social network structure of empirical networks, mainly their homophily and small-world character, and individual perceptions in networks. The family of random geometric graphs constitutes the third strand of literature as a promising methodological choice in Section~\ref{sec:Model}.

\subsection{Stylised Facts on Inequality Perceptions and Middle Class Bias}
The empirical literature has identified four particular stylised facts for any theory of perceived inequality to be evaluated against: $(i)$ Irrespective of their objective status, all individuals perceive themself to be in the middle of the social hierarchy \citep{Kelley1995, Evans2004}; $(ii)$ as an immediate corollary of  $(i)$, poor individuals overestimate their social position, rich individuals tend to underestimate it \citep{Knell2020}; $(iii)$ poor individuals tend to perceive inequality to be higher and are closer to objective inequality on average \citep{osberg2006, Newman2018} and $(iv)$ the evolution of objective inequality is detached from the evolution of subjective inequality, that is, increases in objective inequality do not necessarily increase perceived inequality \citep{Kenworthy2008,Bartels2008,Gimpelson2018,hvidberg2020}. The ubiquity of misperceptions across states and welfare regimes calls for a common mechanism independent of differences in actual inequality or institutional framework.

\begin{figure}[!h]
	\minipage{0.45\textwidth}
	\includegraphics[width=\linewidth]{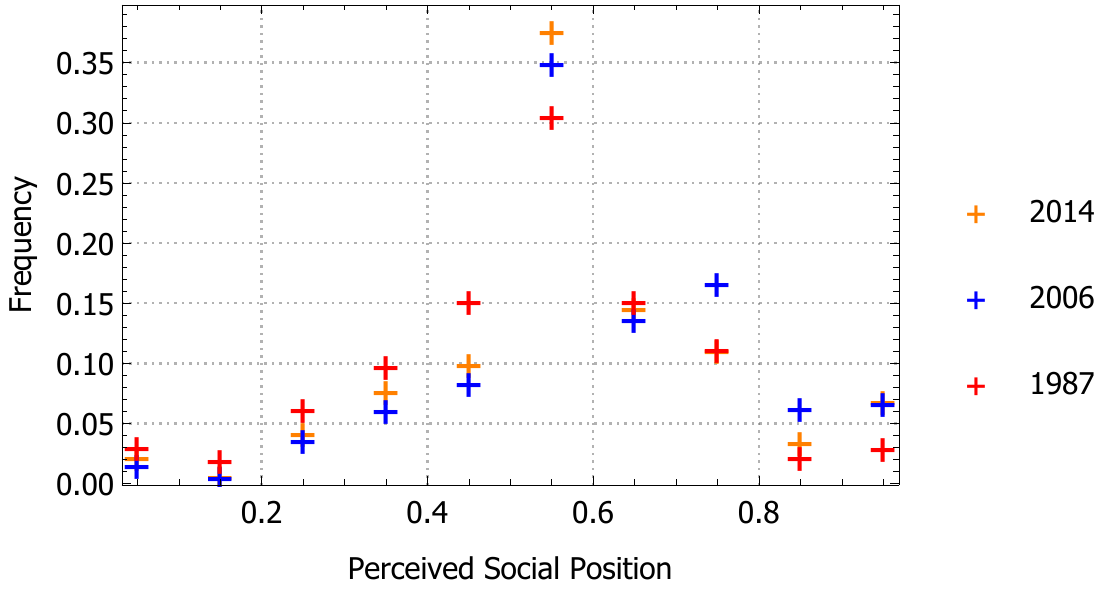}
	\caption{Empirical density of self-reported income decile by US respondents in the ISSP for the waves $1987$, $2006$, and $2014$.}\label{fig:usdec}
	\endminipage\hfill
	\minipage{0.45\textwidth}
	\includegraphics[width=\linewidth]{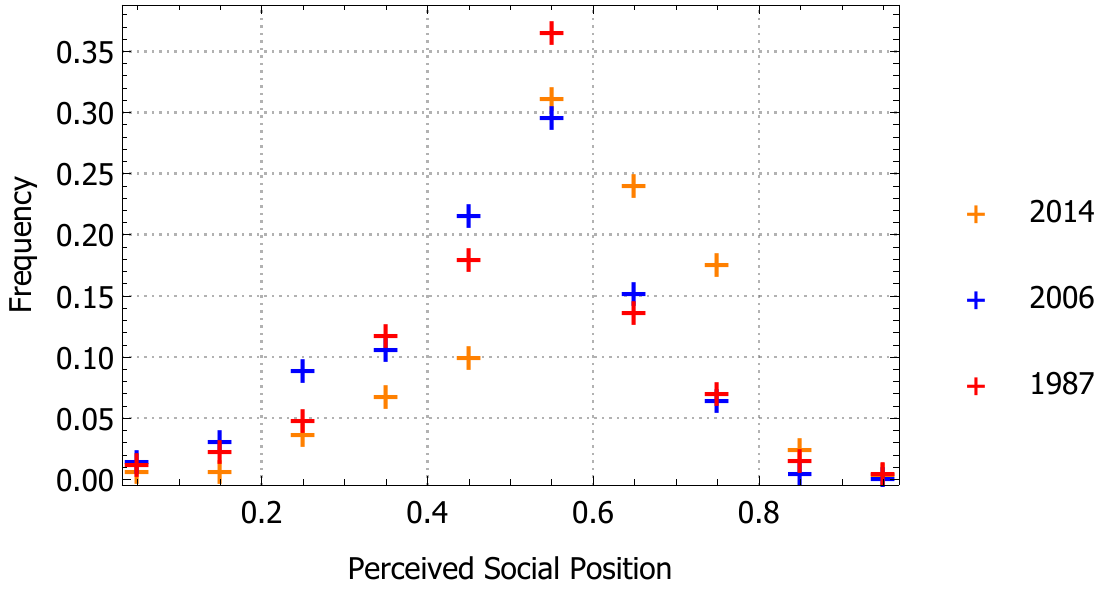}
	\caption{Empirical density of self-reported income decile by German respondents in the ISSP for the waves $1987$, $2006$, and $2014$.}\label{fig:gerdec}
	\endminipage\hfill
	\vspace{0.5cm}
\end{figure}

By way of illustration, we show the empirical frequency of self-perceptions to enable comparisons to the model output in Section~\ref{sec:Results} for stylised fact (i). The data shows the empirical frequency distributions of self-reported income deciles for Germany and the US from the $1987$, $2007$ and $2014$ wave of the ISSP \cite{issp2016}.\footnote{We included exactly these years, since they represent the first, last and median waves of available data.} Typically, Germany is considered to be the epitome of a \emph{coordinated market economy}, while the US represents a \emph{liberal market economy} \citep{hall2001}. By contrast, the qualitatively similar densities for both countries show that the mechanism behind misperceptions should be independent of the specific welfare regime.  Especially the `middle class bias' in the self-perceptions is rather striking. For a representative survey like the ISSP, each decile should, per construction, include exactly $10 \%$ of observations and the frequency densiy should therefore exhibit a uniform density at $0.1$. Instead, the frequencies display a marked peak at the middle categories, i.e., most Germans and most Americans tend to think they are middle class, even though they are objectively not. This finding holds for all considered countries in the ISSP and all considered years, apart from the three periods and two countries we selected for illustrative purposes \citep{Choi2019}.\footnote{The other empirical frequencies are available upon request.} Notice, however, that the qualitative middle class bias manifests itself in quantitatively rather different frequencies throughout time and between countries. We focus on the phenomenon that is common to all considered densities, i.e., the qualitative middle class bias and leave the direct quantitative calibration of our model for further research. We find this middle class bias to be the major driver of our results, implying the other stylised facts (ii) -- (iv) directly but emerging endogenously in our network model. The relevant features of this network are discussed in the subsection below. 

\subsection{Empirical Social Networks}\label{subsec:son}
Empirical networks exhibit ubiquitous and salient features that can serve as stylised facts to guide the validation of proposed theoretical graph formation processes. Probably the most prominent one is the small-world property, indicating that paths between nodes in real-world social networks are unexpectedly short. At the same time, those networks also feature high degrees of clustering. Small-worldiness has obvious implications for any contagion process, be it rumours, diseases or information, where contagion across the whole network happens much faster than our intuition would suggest \citep{watts1999, moore2000, kleinberg2001}. The empirical research has identified small-world features across many different social groups, including friendship networks in schools \citep{weeden2020}, corporate board networks \citep{kogut2001, borgatti2003, davis2003, conyon2006, galaskiewicz2007} and scientific and artistic collaboration \citep{watts1998, newman2001, uzzi2005}. Given this ubiquity, it appears safe to say that a graph-generating process for social networks needs to simultaneously produce low average path lengths and large degrees of clustering to be consistent with this stylised fact. 

We situate our model in the random graph literature, where graph formation happens according to a stochastic process and is not the result of deliberate optimisation. Random graphs have been very successful in replicating structural stylised facts about network topologies, with the Watts-Strogatz model famously able to replicate those small-world properties \citep{watts1998}. Since the graph-generating process is, however, stochastic in nature, it does not feature explicit behavioural microfoundations. Therefore, we extend the purely stochastic notion with a behavioural ingredient, notably, that link-formation is \emph{homophilic}.

First introduced by \cite{lazarsfeld1954}, another salient feature of empirical social networks is \emph{homophily}, the tendency of similar individuals to connect with each other. This tendency is not only an empirical curiosum but has relevant theoretical implications, e.g., for information transmission, where homophilic segregation can severely slow the speed of learning \citep{golub2012} or diminish the attention members of minority groups receive \citep{karimi2018}. The similarity can come in many dimensions such as gender, ethnicity or socio-economic status \citep{Mcpherson2001}. We focus on the latter in the narrow sense of \emph{economic} homophily, which is empirically well-established. One strand of literature focuses on friendship networks in schools and colleges and provides evidence significant homophily according to income or social class \citep{cohen1979, huckfeldt1983, mayer2008, boucher2015, malacarne2017}. 

Even one of the earliest contributions in the field, however, shows that homophily in socio-economic status is not fixed in time and varies with cultural norms and the importance of class distinctions \citep{cohen1979}. In light of this result, it appears unsurprising that we find considerable variation in implied degrees of country-level homophily in our model, perhaps reflecting cultural norms not in the structure but the degree of the graph formation process. Even for a constant degree of (economic) homophily, increasing inequality also increases \emph{segregation} in the population, as then the relative distances in incomes increase and agents becoming more selective in choosing links. For geographical segregation, this effect is empirically documented \citep{reardon2011, chen2012, toth2021}, testifying indirectly to the relevance of homophily as a graph-generating feature. Homophily in income also exists in social media friendship networks \citep[for a large sample of Facebook friends]{lewis2012}, where spatial segregation should not confound findings and becomes apparent in the choice of romantic partners, typically under the label of `homogamy' \citep{kalmijn1991,kalmijn2001}.  Finally, a very recent contribution by \cite{cepic2020} for a representative sample of Croatian adults finds evidence for homophilic tie formation according to social class and income, with however considerable variability in cross-class ties, hinting at possible confounding factors we aim to capture with a parsimonious stochastic process. Importantly, \cite{cepic2020} show that there is also strong variation in cross-class ties, though, which we show might be crucial for unbiased individual inference. A major predictor is there shown to be cross-class sociability is political participation which would, in our model, amount to the prediction that politically active individuals are also better informed on average which is indeed perfectly in line with the evidence \citep[for a recent survey]{campbell2013}.

\subsection{Belief Formation in Networks}
%% Evtl. die Subsection rausnehmen und mit der letzten in einer Überschrift zusammenpacken. Also nichts am Text ändern, nur die Überschriften mergen.
The literature on belief formation itself appears to be much more scarce than the literature on the effects of perceptions and misperceptions. While very different in detail, the two currently dominant theories of public opinion formation suggest that the beliefs an individual holds are in a broad sense averages over the idiosyncratic messages they receive \citep{zaller1992,lodge1995}.\footnote{Cf. \cite{stevenson2013} for a summary on those views.} This literature has focused on the specific `averaging' individuals use to process their information sets. Regarding perception formation about societal inequality, there exist to the best of our knowledge only two analytical models so far, namely \cite{Knell2020} and \cite{iacono2021}. Both derive biased perceptions under very restrictive assumptions, though, and need to impose some kind of `biased averaging'. \cite{Knell2020} assume that agents form subjective income densities over the whole support of possible income levels but let these densities be self-centered such that the mode of the distribution corresponds to their respective own incomes. While they partially succeed in replicating the four stylised facts on inequality perceptions at least qualitatively, their assumption essentially imposes the middle-class bias of stylised fact (i) \emph{ex ante} and not as an emergent outcome from first principles. \cite{iacono2021} also impose potentially biased perceptions and assume that agents only observe their own incomes as well as the minimum and maximum incomes. They continue to derive several important results on voting behaviour and show that information treatments on inequality might alter preferences for redistribution, therefore qualitatively replicating this stylised fact from the empirical literature. While the notion of local information sets appears appealing, using only the minimum, maximum and own income strikes us as unnecessarily artificial and implausible. Most importantly, being static, both models fail to make sense of the - arguably most relevant - stylised fact (iv), i.e., inequality perceptions being very persistent and not responding to changes in actual inequality. Our model is featuring this persistence by exploiting the fact that rising inequality also features rising segregation, as is also empirically established for geographical segregation (cf. the evidence discussed in subsection~\ref{subsec:son}).

We develop on the notion that information is local but assume unbiased processing with skewed information sets as implied by the well-documented homophilic social network formation on which we expand in Section~\ref{sec:Model}. The psychological literature on `social comparison theory' \citep{festinger1954} supports the notion that individual self-perceptions are much more responsive to local knowledge about small groups than to information about aggregates, e.g., knowing the population average \citep{buckingham2002, zell2009, alicke2010}. Thus, belief formation about inequality appears to be indeed primarily based on local knowledge. This does not imply, however, that informational treatments in the form of reported averages do not change beliefs at all. Providing information about the actual degree of inequality seems to exhibit a significant effect on redistributive preferences for Argentina, Sweden and the US \citep{Cruces2013, mccall2017, Karadja2017}, with however small and insignificant effects for Germany \citep{engelhardt2018}. Finally, two recent studies for the whole of Europe and Denmark separately demonstrate that individuals indeed tend to know the income levels of their immediate friends and family rather well, with non-negligible effects on inequality and fairness perceptions as well as perceived social positions \citep{clark2010,hvidberg2020}.
 
Apart from this indirect evidence from informational treatments, there also exist several studies that measure the impact of local exposure to inequality on perceptions and redistributive preferences directly, therefore offering also direct evidence for our proposed mechanism.\footnote{We thank an anonymous reviewer for this hint and the helpful references.} \cite{thal2017} demonstrates using a large-scale survey (Soul of the Community (SOTC)) that affluent Americans' perception of social conditions is largely based on extrapolation from their own neighbourhood, as the affluent within homogeneous and isolated neighbourhoods perceive social disparities to be significantly less severe. \cite{dawtry2015} find robust evidence for the US and New Zealand that individuals base their estimate of average societal income (and other quantiles) largely on their immediate subjective experience or `social sample'. This leads to differences in fairness perceptions and redistributive preferences, whenever the composition of social circles varies by income, as homophily strongly indicates. \cite{kraus2017} finally demonstrate with respect to racial economic inequality that the homogeneity of the immediate social network appears to mask racial inequities, therefore also testifying to the relevance of immediate lived experience for perception formation. 

%komplett neu, allerdings gerne ausbauen :)

In a series of articles close in spirit to our approach, \cite{chiang2011,chiang2015,chiang2015b} exploits this notion and shows experimentally and computationally that individuals base their beliefs about inequality on local perceptions within referent networks and that income homophily has a potentially strong effect on those perceptions. While his approach is exploratory and does not account for the outlined stylised facts on inequality perceptions and empirical social networks, we provide a tractable model, readily calibrated with regards to those phenomena that is introduced below.

\section{Model}\label{sec:Model}
%This section provides a content-oriented presentation; for technical details see the commented model, which is appended electronically/publicly available at GitHub

This section provides a content-oriented presentation; a technical description following the ODD protocol is avaliable upon request. The model consists of three distinct phases run in sequential order:
\begin{enumerate}
	\item Agent initialisation and income allocation
	\item Network formation
	\item Gini perception and network evaluation
\end{enumerate}

Each phase runs only once and phases one and two build the structure which phase three then analyses. This sequence implies that during network generation, agents adapt to others' income level. However, there is no reaction to others' linking behaviour or perception and, thus, the model does not feature interaction in a narrow sense. Moreover, in the model, an agent's social contacts depend on their income. We choose this direction of causality for technical reasons and because it seems empirically likely (cf. Section \ref{sec:Related Literature}). Nevertheless, our process scheduling would also be consistent with the opposite direction of causality or positive feedback effects between income and social contacts.

The model is designed that way because it focuses entirely on income perceptions given defined income distributions and network structures. Hence, both an agent's income and their social contacts remain constant for the evaluated time frame or, put differently, that the simulation outcome is a snapshot of a certain point in time.

\subsection{Agent Initialisation and Income Allocation}
%ACHTUNG: GGF. AUCH LOGNNORMAL NOCH REINSCHREIBEN, WENN WIR DIE NOCH NUTZEN....
There are $1,000$ agents in the model; each agent draws their income from an exponential distribution with a mean of $\lambda = 1$. Such a distribution normalises the empirical observed (pre-tax or market) income distributions in various industrialised countries for the vast majority of individuals \citep{dragulescu2001,silva2004,tao2019}. Thus, one can understand the model population as constituting a representative sample of empirical populations of these countries. The upper tail of $1$ to $5$ \% of the income distributions empirically follows a Pareto law \citep{silva2004}. We deliberately choose to exclude this small minority from our model, since their population size would induce another degree of freedom in our model and we want to demonstrate that segregation is indeed endogenous and not driven by differences in actual income regime. We use an identical, pre-validated exponential distribution for all Monte Carlo runs and also all levels of homophily to ensure comparability between simulation runs. Agents store their true income decile for evaluation purposes, too.

\subsection{Network Formation}
Each agent draws five other agents to link to. Like for real-world networks, links are therefore created \emph{by agents}, not imposed on them. The number of five link choices is also empirically validated, as humans tend to only know the income of close friends or family \citep{clark2010,hvidberg2020}, with typically only five individuals at this closest layer of emotional connection \citep{zhou2005, hamilton2007, Maccaron2016}.\footnote{For a recent review on the large literature on `Dunbar's number', cf. the first section of \cite{Maccaron2016}. However, to the best of our knowledge, there are no empirical studies that specify whom this closest layer consists of.} The relative weight in the draws are a function of the homophily strength and the respective income levels. Thereby, agent $j$'s weight in agent $i$'s draw is denoted by $w_{ij}$ and determined as follows: 

\begin{align}
	w_{ij}&=\frac{1}{\exp[\rho~|I_{j} - I_{i}|]}\label{networkformation}
\end{align}
$I$ denotes the income of an agent, and $\rho \in \mathbb{R}^+$ denotes the homophily strength in income selection, externally set, and identical for all agents. $\rho = 0$ represents a random graph, and for an increasing positive value of $\rho$, an agent becomes ever more likely to pick link-neighbours with incomes being closer to their own. The exponential character of the link function ensures that those others with are large income difference become unlikely picks even at low homophily strengths.  

The choice of an exponential weighting function might seem arbitrary but upon closer inspection, we find that translated into the probability of $i$ choosing $j$, this weighting is equivalent to the discrete choice approach developed and popularised by \cite{manski1981}. The homophily parameter $\rho \in (0,\infty)$ is then simply the intensity of choice parameter. To translate weights into probabilities, we normalise by all weights for all agents, i.e.,

\begin{align}
	p_{ij} &= \frac{\exp[-\rho \cdot |I_{j} - I_{i}|]}{\sum_{k \in M \setminus i}\exp[-\rho \cdot |I_{k} - I_{i}|},\label{discretechoice}
\end{align}
	with $M \setminus i$ as the set of all agents except $i$ with size $N - 1$.\footnote{Note that this is, strictly speaking, only the probability of the first choice of agent $i$, since we consider drawing without replacement,and does not account for the possibility that other agents already link to the agent in question, in contrast to our algorithm. Since the number of agents is rather large, the effect appears to cancel out in the aggregate, though, as also our simulation results in subsection~\ref{subsec:seg} indicate.} This formulation in~\eqref{discretechoice} has a rather intuitive interpretation, with $\rho = 0$ implying equiprobable picks with $p_{ij} = 1/(N-1), \forall j \in M \setminus i,$ and thus indeed a random graph, while $\rho \to \infty$ implies that $p$ approaches unity for $j$ with minimum income distance and $0$ for all other $j$. \cite{manski1981} demonstrate that the discrete choice rule above emerges naturally from random utility theory, i.e., agents maximise utility and utility can be decomposed into an observable and unobservable component. In our case, the observable component the agents minimise would be the income differences, with the unobservable part being all the attributes from which our agent in question would benefit due to their social connection. This appears to be rather intuitive, since of course income differences might be a rather salient characteristic and thus observable, while the utility from social connections might in some cases plausibly exceed the one derived from merely a good fit or small social distance.\footnote{Notice, however, that the derivation of the above choice rule crucially depends on the axiom of Independence of Irrelevant Alternatives (IIA) \citep{luce1977}, i.e., the probability of choosing between $j$ and $k$ being independent of the probability of choosing $l$. IIA might be a good first-order approximation for homophilic choice but in friendship networks, knowing one agent $j$ might indeed increase the likelihood of knowing another agent $l$ that is friends with $j$. It might thus prove interesting to extend and generalise the above choice rule to examine the effects on the network topology in further research.} In this sense, the weighting function in eq.~\eqref{networkformation} is plausibly microfounded in a utility-maximising framework and can now be considered the workhorse choice rule in behavioural macroeconomics \citep{franke2017}. \cite{franke2017} also survey evidence from a several lab experiments in different macroeconomic contexts that discrete choice is indeed consistent with the data, while \cite{anufriev2012} and \cite{anufriev2018} provide laboratory evidence for the discrete choice approach for financial markets. However, there might of course be other potential choice mechanisms that could provide avenues for further research on network generation that can be readily included within our proposed flexible RGG framework. 
%komplett neu
\begin{figure}[!h]
	\begin{center}
		\minipage{0.75\textwidth}
		\includegraphics[width=\linewidth]{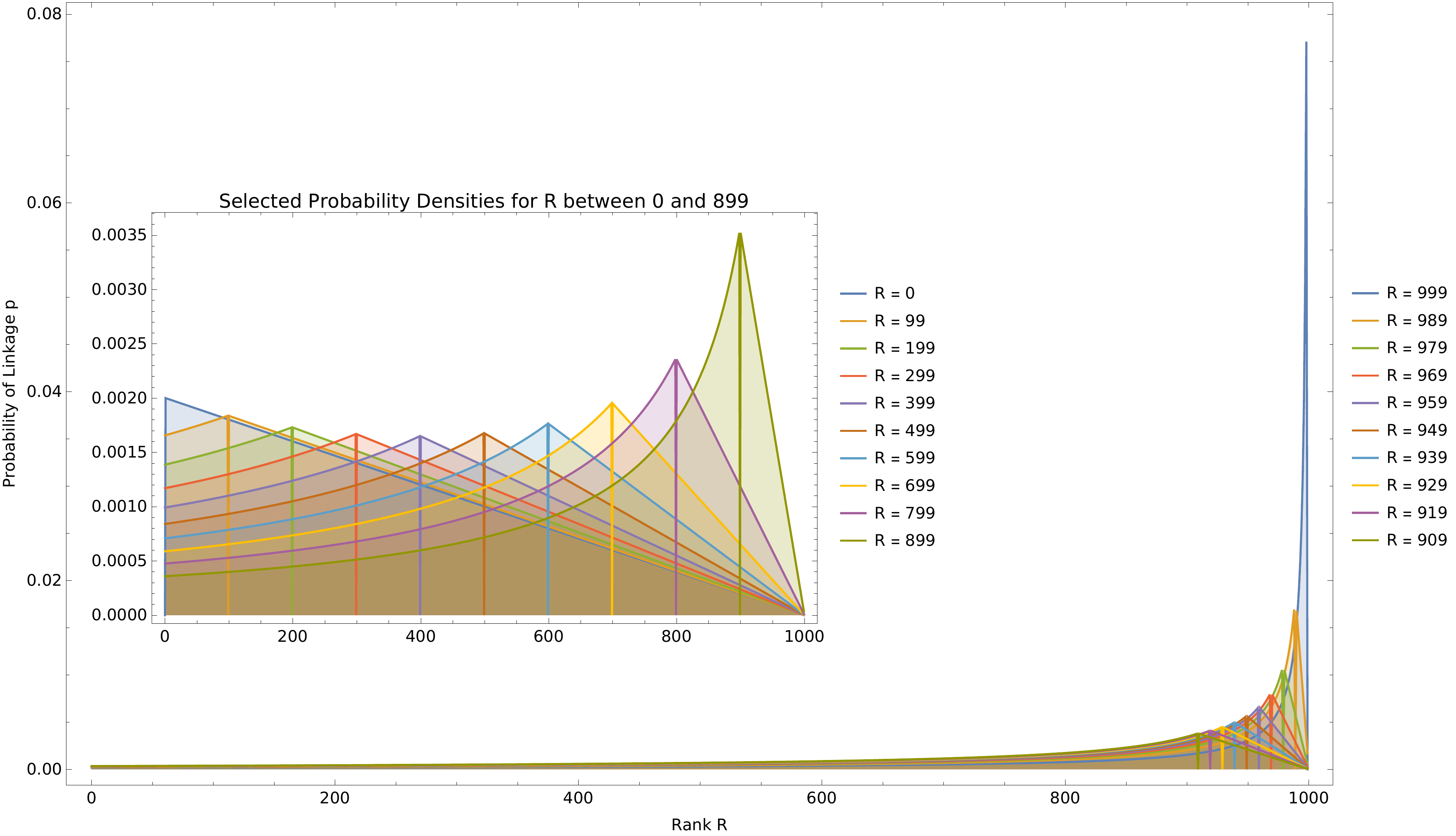}
		\caption{Theoretical PDF of Linkage Probabilities for Ranks $R$ and $\rho = 1$.}\label{fig:pdfrho1}
		\endminipage\\
	\end{center}
	
	\begin{center}
		\minipage{0.75\textwidth}
		\includegraphics[width=\linewidth]{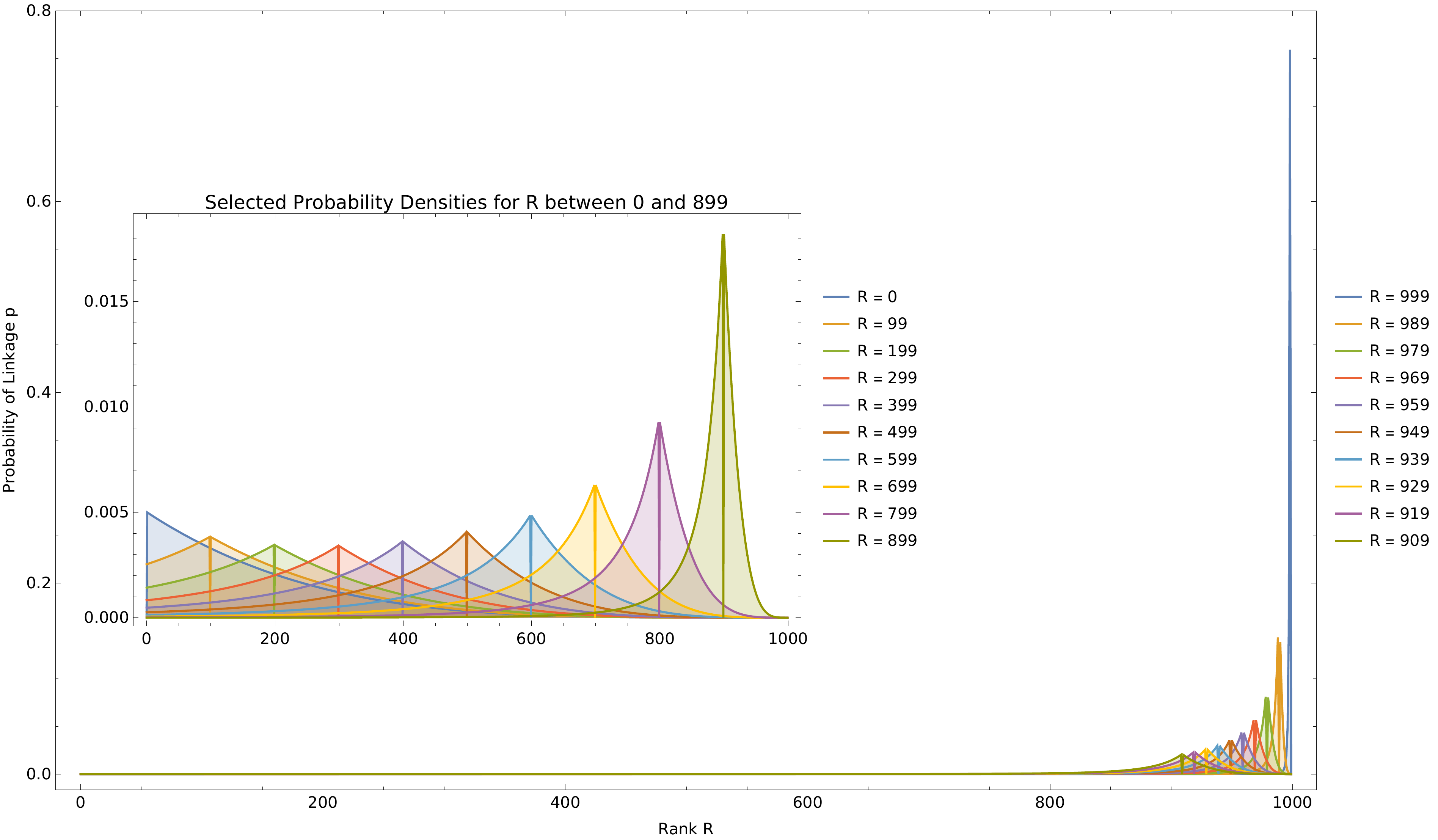}
		\caption{Theoretical PDF of Linkage Probabilities for Ranks $R$ and $\rho = 4$.}\label{fig:pdfrho4}
		\endminipage
		\vspace{0.5cm}
	\end{center}
	
	\footnotesize{\emph{Note:} The Figures plot the Probability Density Functions (PDFs) of a node with a given income rank for linkage with another node for the whole support of income ranks. The above Figure assumes a homophily strength $\rho = 1$, whereas the below Figure assumes $\rho = 4$.}
\end{figure}

Figures~\ref{fig:pdfrho1} and \ref{fig:pdfrho4} illustrate the linkage probabilities implied by the weighted draw based on the exponentially distributed income levels. As can be seen, the decay within the left tail is always more rapid than for the right tail, indicating differences in the `selectivity' above or below a relative position. We understand `selectivity' according to rank as the effect a decrease in income rank distance of one agent to another has on the linkage probability between them. Consequently, the local maxima of individual linkage probability densities exhibit a bi-modal shape with peaks at the highest and lowest rank but are also heavily skewed to the left, i.e., agents with the high incomes are most selective in their link picks. General selectivity increases with $\rho$. Notice also that largest income ranks are extremely selective in all scenarios, in some cases in some cases exceeding linkage probabilities of incomes close to the median by more than two orders of magnitude in linkage probabilities.

An anonymous reviewer pointed out that individuals may form links based on relative rather than absolute income differences: For lower incomes, a given absolute gap in units of currency may mean two entirely different lived experiences, while people with high incomes may hardly notice the same absolute gap. To represent this in the linkage function in eq.~\eqref{networkformation}, one must simply replace $I_{i}$ by $\ln(I_{i})$ and $I_{j}$ by $\ln(I_{j})$. \ref{app:logincome} analyses this transformation of scale in detail. The altered argument in the choice function is equivalent to assuming that agents aim to minimise the percentage difference in incomes and is, therefore, a natural extension to capture potentially scale-dependent tie-formation along the lines discussed above \citep{tornqvist1985}.

Our findings from Section~\ref{sec:Results} regarding self-perceptions and aggregate inequality perceptions prove qualitatively robust; quantitatively, the major findings occur at even lower homophily levels. However, the logarithmised incomes fail to replicate the greater under-estimation of inequality for richer individuals with higher income ranks. The segregation tendency is approximately symmetric for moderate to high homophily strengths, i.e., all agents are approximately equally likely to include agents below and above them in income rank. Since there is hardly any differential behaviour according to income rank, all agents tend to perceive roughly equal levels of inequality according to the (local) Gini, in contrast to the stylised fact (iii). Thus, absolute income differences pose the more strongly validated presentation in the present model framework. Nevertheless, the model invites empirical research into whether income homophily is based on absolute or relative differences - or some combination of both.\footnote{Since only the specification following scale-independent choice is consistent with inequality perceptions varying systematically with income while choice based on logarithmised incomes is not, the existence or non-existence of differential behaviour might help to discriminate between implied choice functions.}

The resulting network for our baseline specification is a member of the family of Random Geometric Graphs \citep{dall2002}, which \cite{talaga2020} showed to reproduce core features of many social networks efficiently. Specifically, we combine the notions of homophily \citep{boguna2004} with pre-setting node degrees \citep{newman2001a,newman2009}.\footnote{Our procedure is related to \cite{karimi2018} who also combine a Preferential-Attachment model with a homophilic ingredient. Crucially, however, their model builds on a binary notion of homophily with only two groups. Our algorithm, in contrast, does not impose any restriction on the target feature and is applicable to attributes potentially defined over the whole positive real half-line.} However, concerning our application, we are able to simplify both approaches by pre-determination of only the global minimum degree, like in Preferential-Attachment networks, and consequently defining relative weights rather than absolute probabilities.

Links are undirected and have identical weights for evaluation purposes. Agents pick their neighbours in random sequential order. If an agent $i$ picks agent $j$ who had themself picked $i$ before that, the already existing link between the two agents remains untouched, but $i$ does not pick another neighbour instead of $j$. Consequently, each agent has at least 5 link-neighbours (i.e. clsoe social contacts with mutual knowledge of income) but may have more.

\subsection{Gini Perception and Network Evaluation}
Agents know about their own income and also their social contacts' incomes. However, they do not possess knowledge about any other agent or  structural features of the whole income distribution. Thus, agents judge income inequality in the population as well as their own income position solely based on themself and their link-neighbours. Besides the agents' perceptions, there is a global assessment of various network parameters in order to validate the model.

Subjective inequality perceptions mirror standard Gini calculation on the level of individual personal networks: Each agent finds the mean of all income differences between themself and each link neighbour and between any two of their link-neighbours and divides this by twice the mean overall income of themselves and all link-neighbours. Then, the overall perceived Gini is simply the arithmetic mean of individual perceptions.

To estimate their income decile, an agent compares the number of link-neighbours having a higher income than the agent themself to the link-neighbours having a lower income than the agent themself.

%subsection zu validation (wie nennen?)
%ERGM/SAOM nicht passend, da Daten fehlen (De Paula, 2020) und nicht analytisch lösbar
% unser Modell zumindest teilweise analytisch lösbar (siehe Appendix) + well-grounded in established behavioural theory (random utility)
% assumptions are analytically convenient but does not imply that this is only generating mechanism: we hope to inspire further research for grounding homophilic attachment in EUT

\subsection{Validation}
The simulation results of our model are in line with our theoretical expectations and we can explain their emergence in terms of the mechanisms sketched in Section~\ref{sec:Results}. Moreover, we carried out sensitivity analyses that revealed no unintended consequences of changes in any relevant model feature like homophily level, number of links or actual income distribution. Thus, we consider the model design and implementation to be internally validated \citep[p. 22]{gilbert2005} as a tool for explaining inequality perceptions in the model population.

Transferring these explanations from the model to the real world requires external validation of our model.  However, there are different accounts of what constitutes an explanation in the first place. The current discussion of the concept of explanations in the philosophy of the social sciences highlights two types of explanations: How-actually and how-possibly explanations, also known as candidate explanations \citep{Epstein1999}. While how-actually explanations aim for identifying the actual mechanism driving the dynamics in a specific case, how-possibly-explanations provide mechanisms that could possibly bring about the explanandum in question \citep{reutlinger2018}; they enquire for mechanisms that potentially cause the observed phenomenon. In case of epistemically possible how-possibly explanations, these mechanisms are in line with the knowledge about the real world \citep{grune2021}.

Our model yields an epistemically possible how-possibly explanation of inequality perception because it ``produces quantitative agreement with empirical macrostructures, as established through on-board statistical estimation routines'' and also ``quantitative agreement with empirical microstructures, as determined from cross-sectional and longitudinal analysis of the agent population'' \citep{barde2017}: Simulation outputs of a societal structure close to a small-world one with self-segregation of highest-income agents and severe underestimation of the income Gini across income levels mirror the corresponding empirical findings.

Following the suggestion by \cite{fagiolo2019}, we use empirical micro-data to calibrate the model. Namely, it relies on an exponential income distribution that characterises industrialised countries. Furthermore, the  extent of agents' closest layer of interaction (`Dunbar's number') that means mutual knowledge of income, their linking behaviour, and individual perception formation follows rules that are theoretically established in rational choice theory but also empirically grounded in the referenced lab experiments and surveys. The exponential weighting function from the discrete choice framework is also analytically convenient and lets us represent the probability densities of ties in closed form. This allows us to e.g. demonstrate conclusively that the combination of discrete choice in graph formation and an exponential income distribution leads to the endogenous emergence of echo chambers for top-income earners whose isolation increases in the intensity of choice $\rho$.\footnote{
	This combination of analytical convenience that leads to internal validity and empirical plausibility that affirms external validtity is also one of the reasons why we deliberately choose not to use an Exponential Random Graph (ERGM) or Stochastic Actor Oriented Modelling (SAOM) framework  \citep[for a recent survey]{snijders2011} but situate our model in the RGG framework: Firstly, the application of these types of models would require merging relational data with the socioeconomic status of the respective agents which is rarely achieved in practice, as \cite{depaula2017} notes. In our case, the problem of data availability is compounded by the fact that we require the graph data not only to report all social ties but also to identify the closest layer of emotional connection. Only there we can reasonably expect agents to exactly observe incomes as is required by our model mechanism. We are currently not aware of any dataset fulfilling these constraints but welcome any empirical attempt in this direction, as the external validity of our proposed model mechanism can ultimately only be established empirically. Secondly and more importantly, the estimated coefficient estimates and tie-level probability densities from ERGMs and SAOMs are purely phenomenological and need to be simulated by Monte Carlo techniques, while we are able to express them analytically and thus precisely determine the effect of our model parameters. We thank an anonymous reviewer for pointing us to ERGMs.} 

This empirical input calibration and output validation jointly guarantee resemblance \citep{maeki2009} between our model and the real world. We develop a specific parallel reality \citep{sugden2009} that features generating mechanisms for empirical findings in our reality, and hence our results present a candidate explanation for the stylised empirical facts. There may be different, more adequate, parallel realities featuring either these or even better mechanisms, despite to the best of our knowledge there being no existing models that fulfil these characteristics. Overall, the following section presents an epistemically possible how-possibly explanation of inequality underestimation that ``constitutes epistemic progress on the way towards HAEs [how-actually explanations, A/N]'' \citep{grune2021} of the phenomenon. The model simultaneously features technical verification and external validation based on input and output measures. \cite{grabner2018} considers this combination desirable albeit rarely possible for model development.
Since our model features a range of proposed micro-mechanisms (e.g., on endogenously evolving segregation, cf. subsection~\ref{subsec:perc}), we also hope to inform empirical research to further examine their external validity.

\section{Results}\label{sec:Results}
The homophilic graph model will be evaluated against the five stylised facts outlined earlier. As we have shown in Section~\ref{sec:Model}, we only require the homophily strength parameter $\rho \in \mathbb{R}_0^{+}$, the number of links each node chooses $C$ and the income distribution as inputs for initialisation. Since link formation is stochastic, we run the graph formation routine $100$ times and report model averages, if not otherwise indicated. Most of the results are obtained with initialisation by the same set of incomes generated from an exponential distribution with location parameter $\lambda = 1$ and $1,000$ observations for $ C = 5$ choices of link-neighbours each agent undertakes to make results comparable for variation in $\rho$. The overall Gini coefficient for these $1,000$ randomly generated income levels is with $G \approx 0.50701$ within $1.5$ \% deviation from the theoretical Gini of $G = 0.5$, indicating that the observed effects of $\rho$ are not artefacts of initialisation. Results are also robust for different numbers of links chosen per node, as long as $C \ll N$.  We also evaluated the null model for $\rho = 0$, where we did not find any significant deviations in the mean inequality perceptions and the actual overall inequality of $G = 0.5$, testifying to the robustness of our approach.\footnote{The results for the null model as well as for different $C$ are available upon request.}

\subsection{Small-Worldiness}
We use state-of-the-art methods to test for the existence of small-world features against an appropriate network null model, here an Erd{\H{o}}s-R{\'e}nyi (ER) graph with the corresponding number of nodes and mean degree first introduced by \cite{erdos1960}. ER graphs appear to be the correct null model for two reasons: Firstly, they are a particular case of our model with $\rho = 0$, i.e. without homophily. Hence, the procedure allows isolating the impact of homophily and examining whether the model indeed tends to yield `smaller worlds' for homophilic formation in the precise sense outlined below. Secondly, we can establish an exact one-to-one correspondence between a graph generated by our model and the ER model, as ER graphs only require the number of nodes and a linkage probability for initialisation that is fully determined by the mean degree of the correspondent network. Other prominent generating models such as Watts-Strogatz graphs have additional degrees of freedom like the `rewiring probability' without clear correspondence to our model.

We construct three summary metrics to test our model against, as introduced by \cite{humphries2008}. Firstly, $\Lambda$ measures the deviation in average path lenghts $L$, that is,

\begin{align}
	\Lambda_i &:= \frac{L_i}{L_i^{ER}},
\end{align}
where $L_i$ is the average path length of network $i$ with $L_i^{ER}$ as the average path length of a correspondent ER graph with equivalent number of nodes and mean degree. `Small-worldiness' requires $\Lambda \approx 1$, as our network should not deviate too much from the random benchmark that indeed features short paths. $E [L_i^{ER} ] = (\log[N] - \gamma)/(\log[k])) + 1/2$ with $\gamma$ as Euler's constant, $N$ as the number of nodes and $k$ as the average degree can be analytically  derived  which we use in our calculation \citep{fronczak2004}.

Secondly, we also require a high clustering coefficient which an ER graph cannot generate. The deviation in the clustering coefficients $\Gamma$ is defined as

\begin{align}
	\Gamma_i &:= \frac{C_i}{C_i^{ER}},
\end{align}
with $C_i$ as the clustering coefficient of graph $i$ and $C_i^{ER}$ as the clustering coefficient of the corresponding ER graph. Here, again, analytical results are available which we utilise, mainly that $E[C_i^{ER}] = k/N$ with again $k$ as the average degree and $N$ the number of nodes \citep{watts1999}. Since ER graphs typically do not exhibit clustering, we require here that $\Gamma_i > 1 $ for a small-world to be present. 

Finally, we use a summary measure $\Phi$ introduced by \cite{humphries2008}. We define $\Phi$ as

\begin{align}
	\Phi_i &:=  \frac{C_i}{C_i^{ER}}/ \frac{L_i}{L_i^{ER}}  = \frac{\Gamma_i}{\Lambda_i}.
\end{align}
\cite{humphries2008} show that $\Phi_i$ features desirable statistical properties when confronted with the conventional Watts-Strogatz model for graph formation and shows a unique maximum between the extreme cases of a random network and an ordered lattice. This is in line with our intuition that small-worldiness results from the interaction of order (in the form of high clustering near the lattice) and randomness (in the form of the random graph featuring low average path lengths), as shown by \cite{watts1998}. We require $\Phi > 1$ for small-worlds. Note that $\Phi > 1$ is an immediate corollary of the two requirements $\Gamma > 1$ and $\Lambda \approx 1$, but $\Phi > 1 $ does not imply the two individual requirements. We call the first sufficient condition `strong small-worldiness' and $\Phi > 1$ with a violation of either  $\Gamma > 1$ or $\Lambda \approx 1$ `weak small-worldiness', where we now only require normalised clustering to increase \textit{faster} than average path lengths.

\begin{figure}[!h]
	\minipage{0.32\textwidth}
	\includegraphics[width=\linewidth]{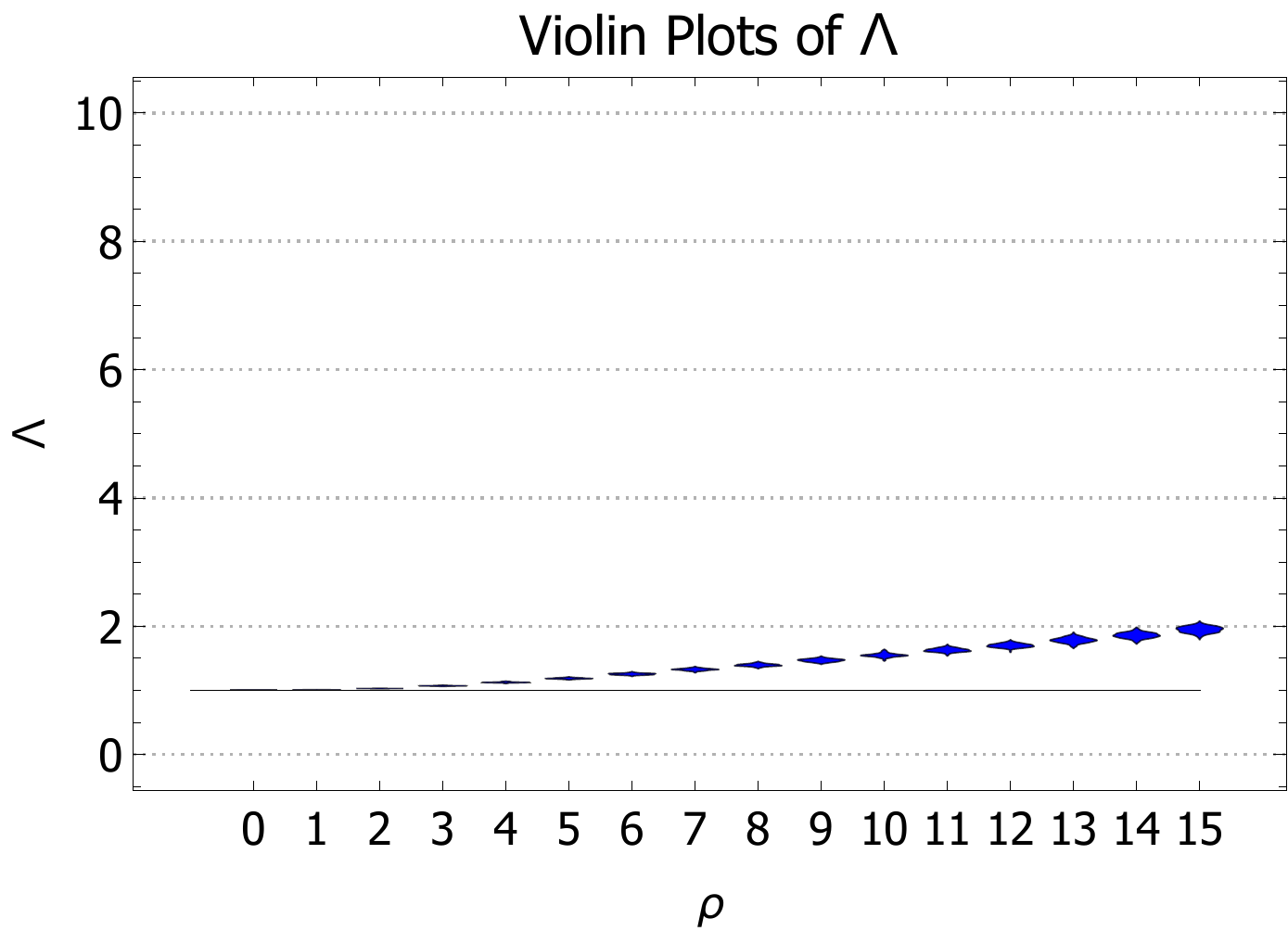}
	\caption{Violin Plots of Normalised Average Path Lengths $\Lambda$ as a function of Homophily Strength $\rho$.}\label{fig:lambda}
	\endminipage\hfill
	\minipage{0.32\textwidth}
	\includegraphics[width=\linewidth]{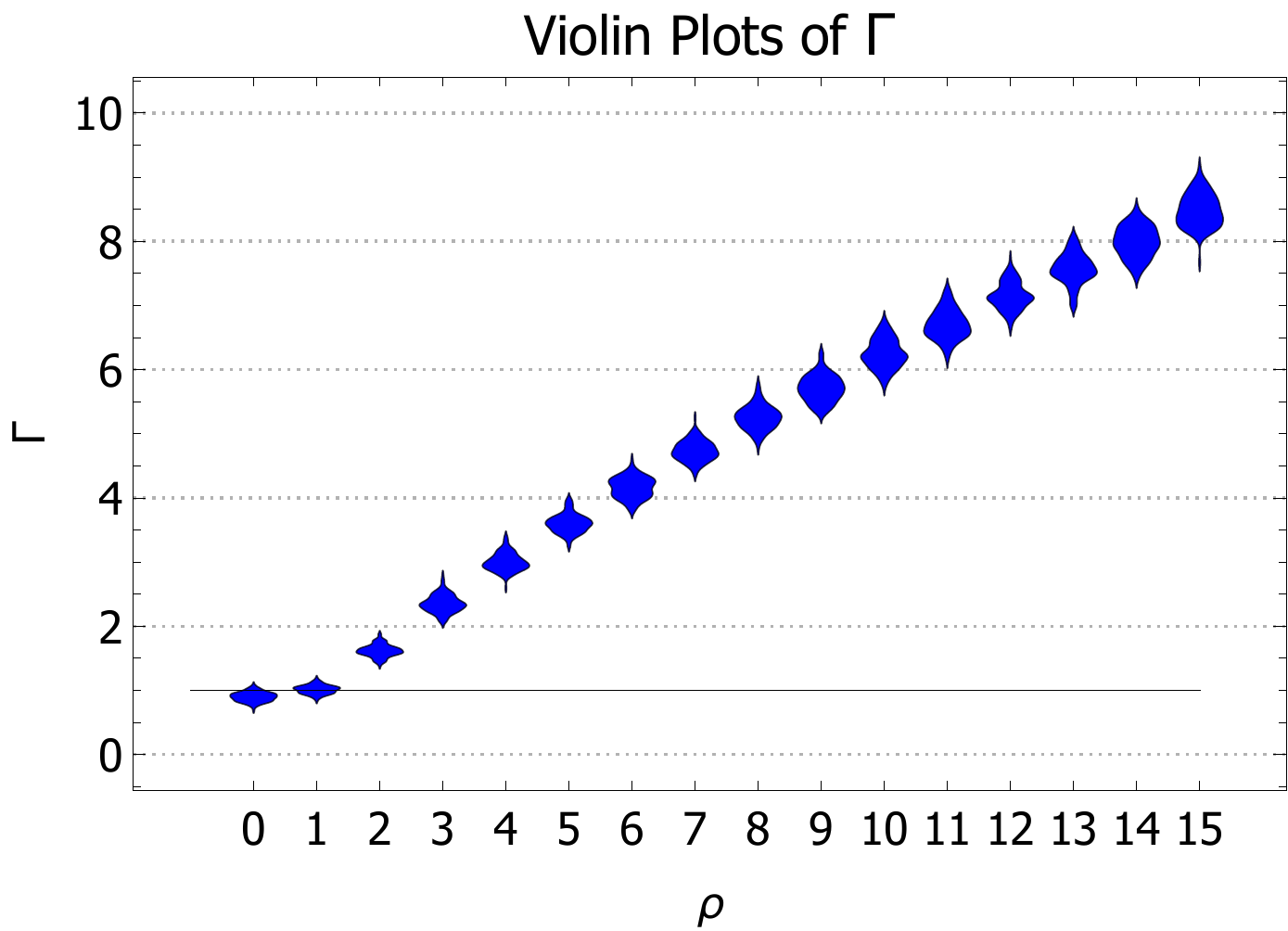}
	\caption{Violin Plots of Normalised Clustering Coefficients $\Gamma$ as a function of Homophily Strength $\rho$.}\label{fig:gamma}
	\endminipage\hfill
	\minipage{0.32\textwidth}%
	\includegraphics[width=\linewidth]{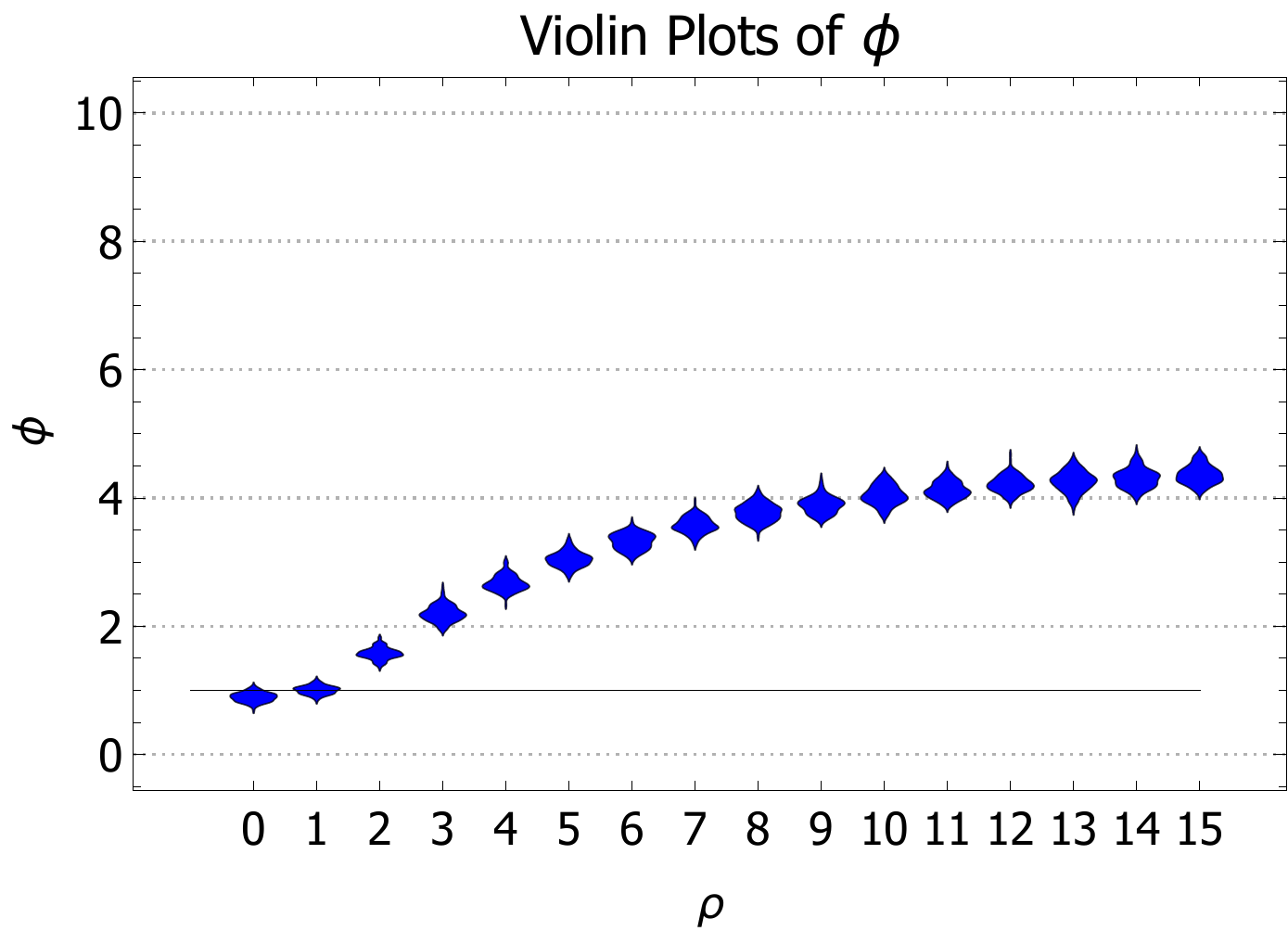}
	\caption{Violin Plots of Small-World Summary Statistic $\Phi$ as a function of Homophily Strength $\rho$.}\label{fig:phi}
	\endminipage
	\vspace{0.5cm}
	
	\footnotesize{\emph{Note:} The Figures report violin plots for the relevant statistics for `small-worldiness'. The average path length is significantly higher than the ER benchmark for all depicted $\rho$, indicating violation of the `strong small-worldiness condition'. Normalised clustering coefficients are for $\rho > 1$ significantly higher than the ER benchmark and increase at a much faster rate than average path lengths, indicating that indeed the `weak small-worldiness' condition is fulfilled.}
\end{figure}

The ER benchmark is nested in our model for $\rho = 0$, as is also readily visible from the fact that both $\Lambda \approx 1$ and $\Gamma \approx 1$ for $\rho = 0$. We indeed find that homophily induces path lengths to grow significantly above this ER benchmark. 
% We speculate that this is a result of the endogenous segregation of the richest decile our model generates, as discussed below in section~tbd. 
Normalised clustering coefficients, however, increase much more rapidly with homophily than average path lengths, demonstrating that our model can achieve relatively high clustering without simultaneously increasing path lengths in the same way. The proposed process thus violates the strong condition but fulfills the weak condition for small-worlds and is therfore broadly in accordance with the topological patterns found in real-world social networks. We note further the symmetry to the canonical Watts-Strogatz approach \citep{watts1998}. While we build on a random network with short average path lengths and interpolate to the desired high clustering through homophily, Watts and Strogatz start from an ordered state with high clustering and approach the random graph benchmark by rewiring to generate shorter average path lengths. Arguably, however, our approach starts from a plausible and empirically well-established behavioural principle in contrast to the purely stochastic process in the Watts-Strogatz world without such behavioural foundations. Besides providing empirical validation, this finding might also point to relatively rapid contagion throughout the homophilic network, be it in the form of rumours or `expenditure cascades'.

\subsection{Perceived Social Hierarchy and Middle Class Bias}
\begin{figure}[!h]
	\minipage{0.24\textwidth}
	\includegraphics[width=\linewidth]{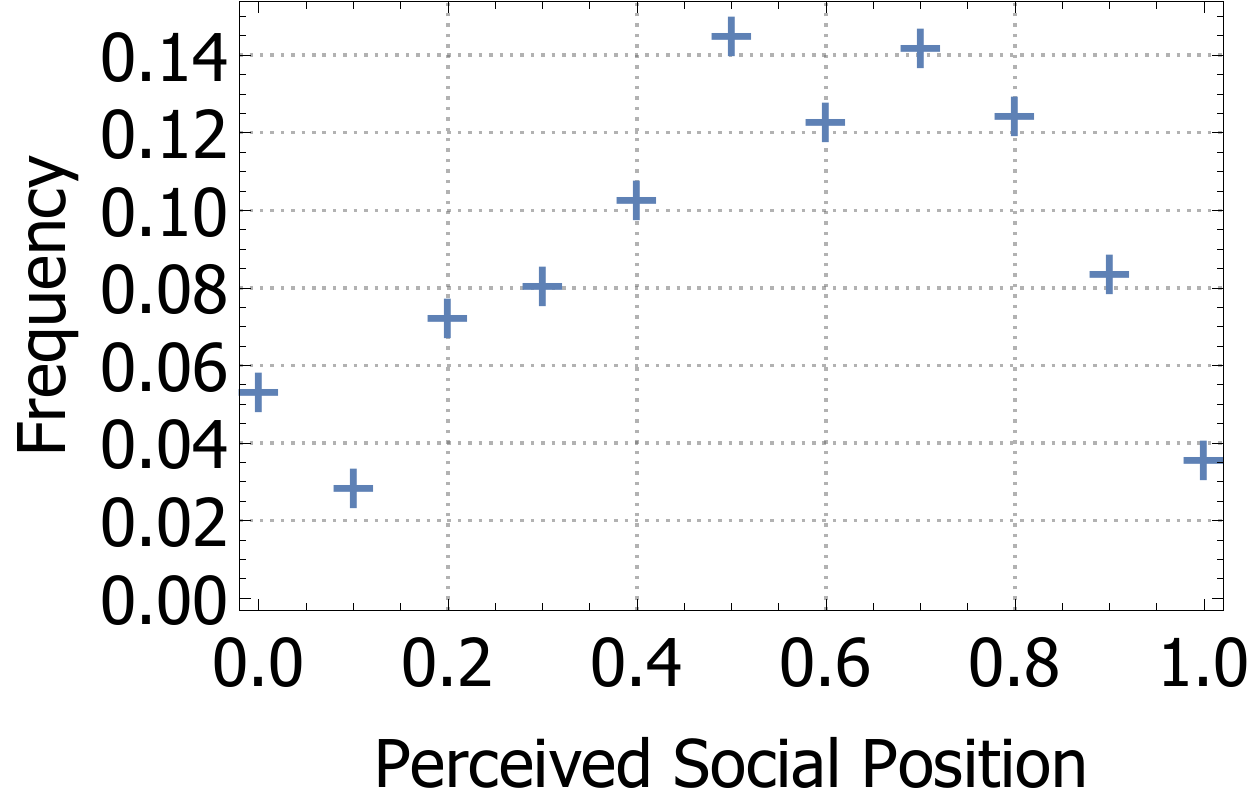}
	\caption{ Density of Perceived Quantiles for $\rho = 1$.}\label{fig:denshs1}
	\endminipage\hfill
	\minipage{0.24\textwidth}
	\includegraphics[width=\linewidth]{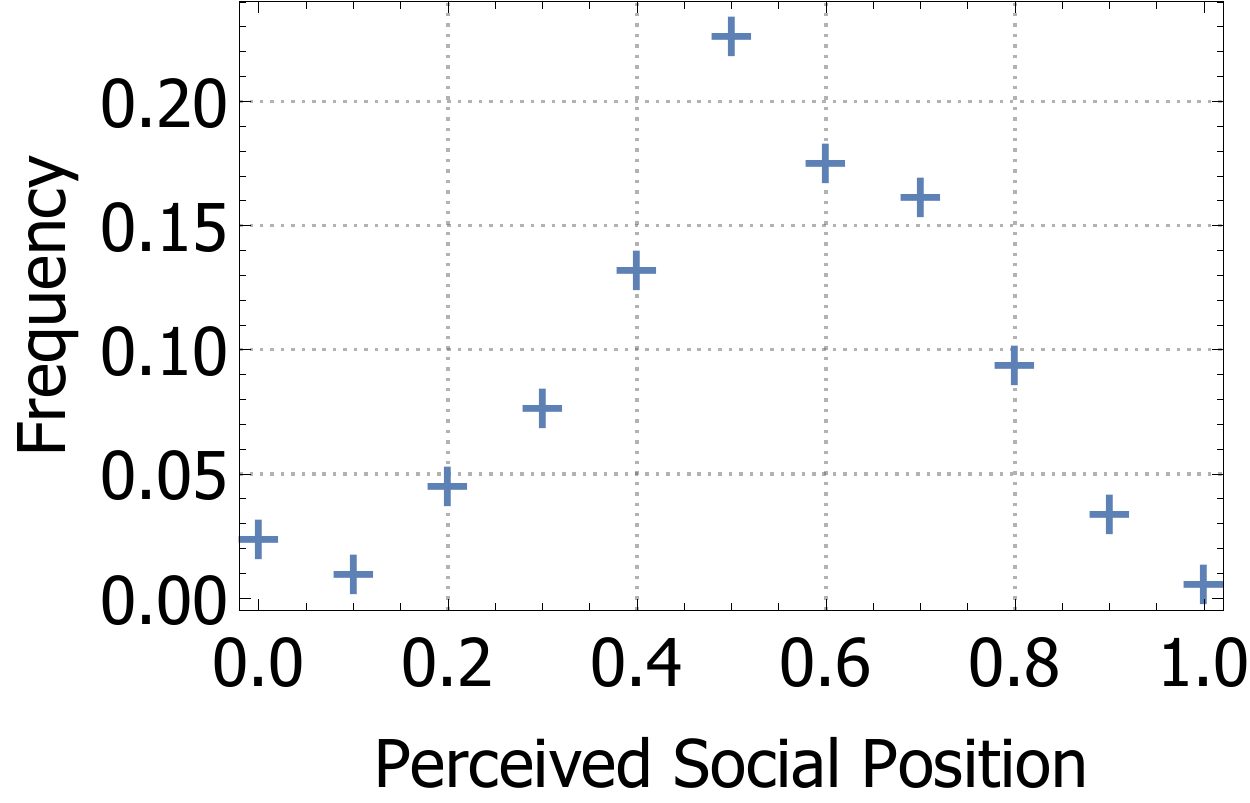}
	\caption{Density of Perceived Quantiles for $\rho = 4$.}\label{fig:denshs4}
	\endminipage\hfill
	\minipage{0.24\textwidth}%
	\includegraphics[width=\linewidth]{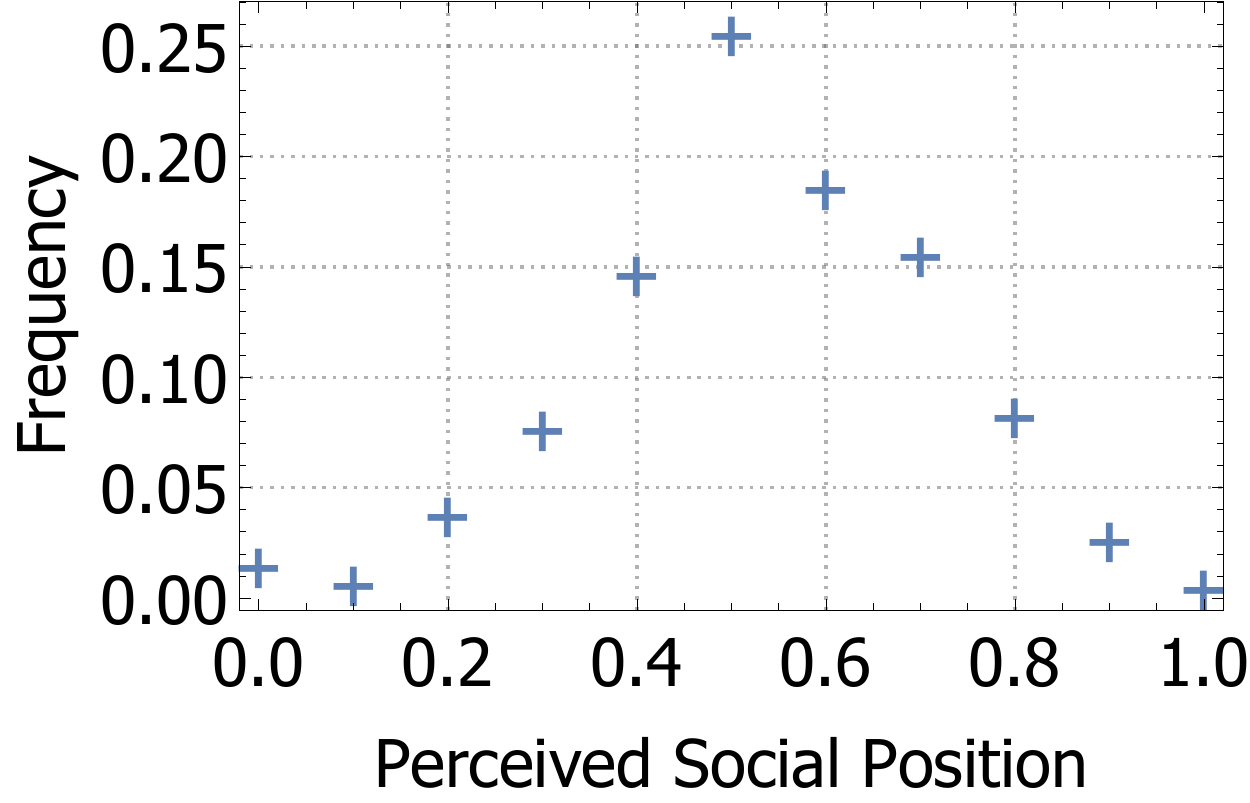}
	\caption{Density of Perceived Quantiles for $\rho = 8$.}\label{fig:denshs8}
	\endminipage\hfill
	\minipage{0.24\textwidth}%
	\includegraphics[width=\linewidth]{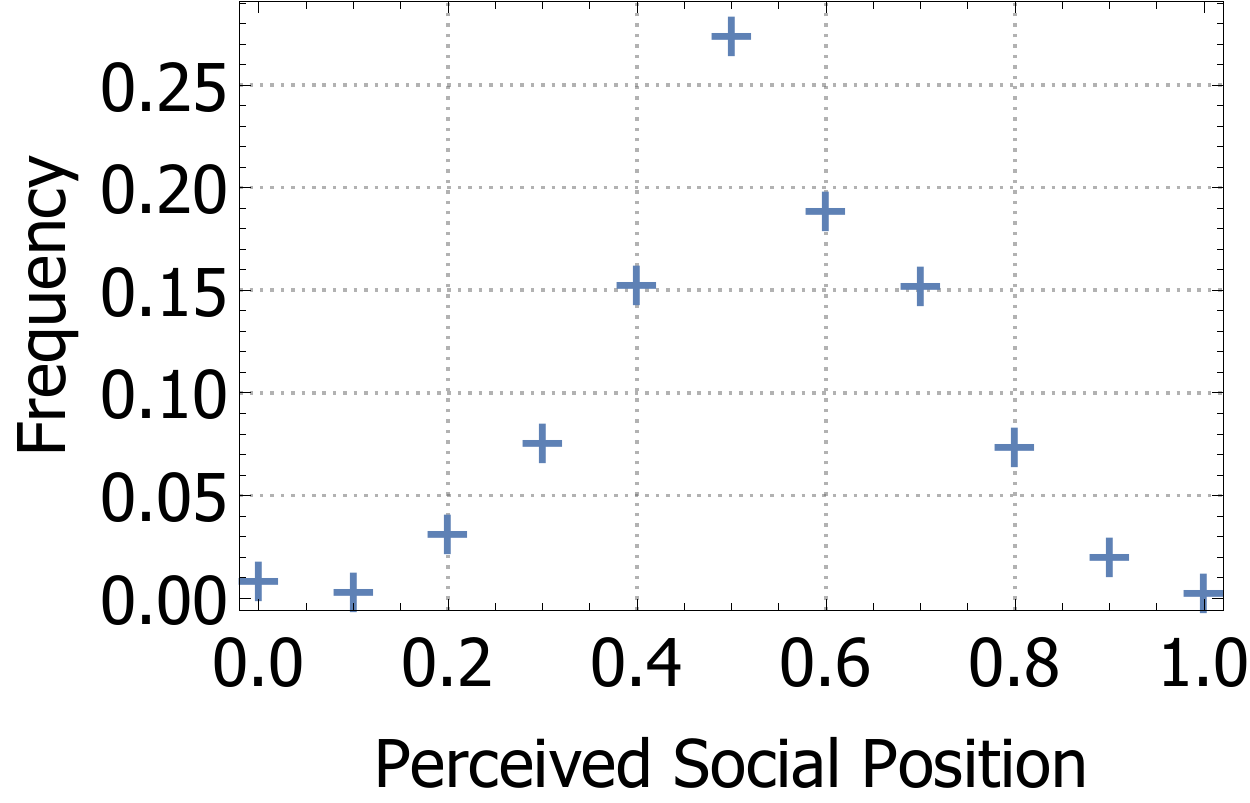}
	\caption{Density of Perceived Quantiles for $\rho = 14$.}\label{fig:denshs14}
	\endminipage
	\vspace{0.5cm}
	
	\footnotesize{\emph{Note:} The Figures report the perceived social positions for $\rho \in \{1;4;8;14\}$ with $10$ bins each. All Figures exhibit significant deviation from the benchmark with equal frequencies. The tendency for individuals to place themselves in the middle of the income hierarchy is, however, only apparent for the middle and right panels, indicating that a homophily strength $\rho$ of $1$ might be too low to replicate the empirically observed tendency. For $\rho = 4$, $8$ and $14$, the densities approximate the empirical densities rather well, though.}
\end{figure}

For unbiased hierarchy perceptions, the reported frequency of perceived social position would coincide with the actual positions. Unbiased perceptions thus entail reported perceived positions of equal frequency, as they coincide with the actual population shares. As we show both analytically in~\ref{app:percdec} and by simulation, perceived social positions for homophilic graph formation are far from the equiprobable benchmark. We find a tendency of the vast majority of individuals to place themselves in the middle of the perceived hierarchy, in line with the empirical evidence. We prove that the tendency exists for all $\rho \in (0,\infty)$. Its strength is a function of $\rho$, though, as we show exemplarily in Figures~\ref{fig:denshs1} to~\ref{fig:denshs14}. The figures plot the empirical densities of income ranks which the respective  the individuals perceive to hold. For $\rho = 1$, the tendency is relatively weak, while for $\rho = 4$, $\rho = 8$ and $\rho = 14$, the densities display a distribution that notably peaks at the centre. In fact, the displayed densities indeed seem to feature all the salient features of the densities of empirical perceived social positions, as shown in \cite{Choi2019} and also in Figures~\ref{fig:usdec} and~\ref{fig:gerdec} for the ISSP data we collect.

%hier noch den Verweis zu den Figures oben eingebaut

Notice that this a necessary \emph{outcome} of homophilic graph formation under very mild and general conditions and based on a well-established utility maximisation framework, in contrast to models that take this tendency as an \emph{assumption}. The latter strand of literature has typically taken a bounded rationality stance on the issue and argued that it is failures in information processing which explain the persistent errors in perceptions of social positioning. Our model replicates stylised fact $(i)$ purely by virtue of the network formation process. In contrast to the literature on bounded rationality, we can hence show that stylised fact $(i)$ is consistent with purely rational actors that form correct beliefs based on their available information, as long as homophilic graph formation constrains their information sets. Our model thus entails very different policy implications to improve self-perceptions. Since information processing is assumed to be correct in our model, information treatments, i.e., increasing the information received from nodes with heterophilic incomes, have mitigating effects on perceptions. Influencing information processing itself, as implied by the established models, is arguably a much harder task for policy.

\begin{figure}[!h]
	\begin{center}
		\includegraphics[width=.8\linewidth]{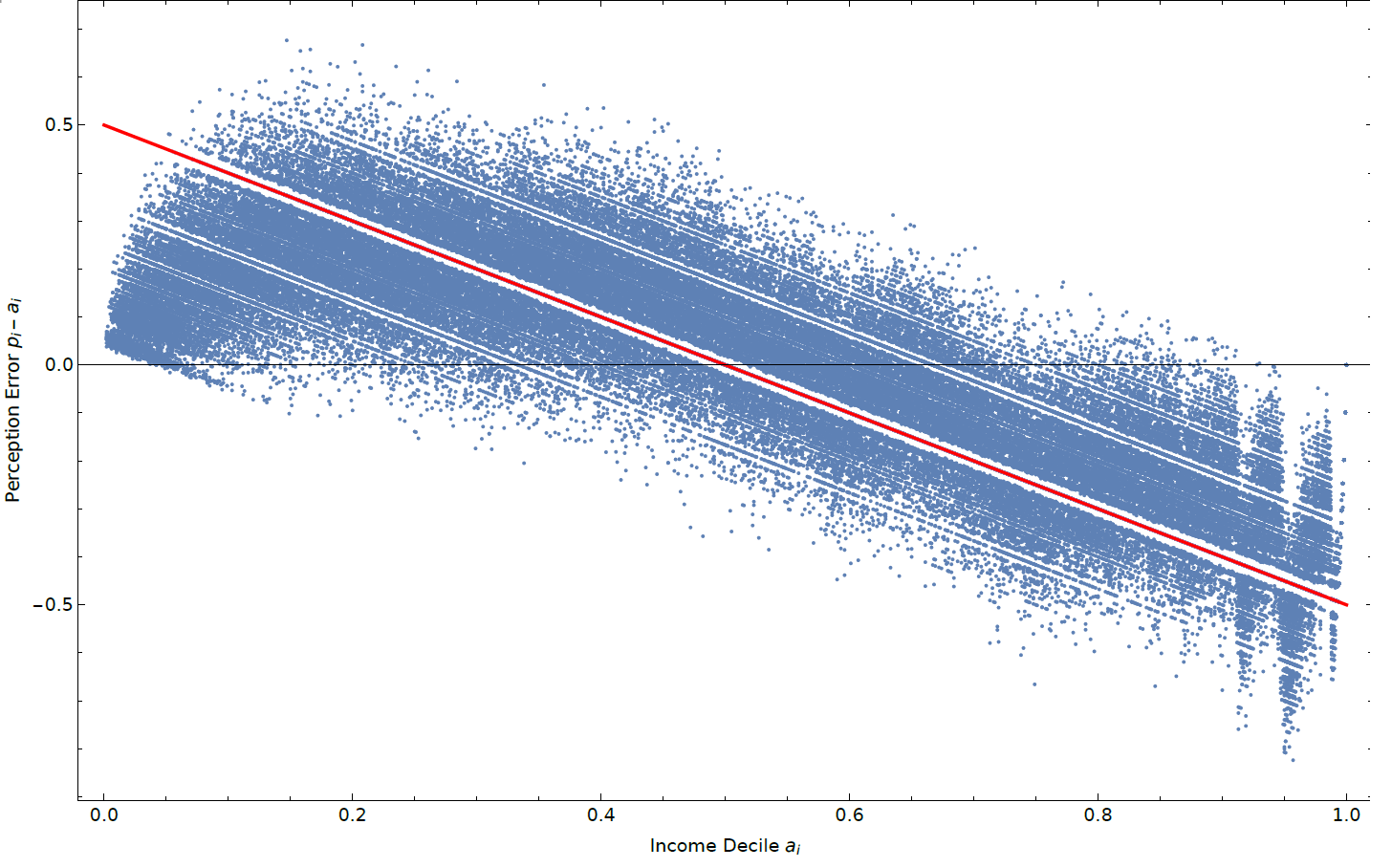}
		\caption{Errors $\epsilon_i$ show the difference between perceived position $q_i$ and actual position $a_i$ for all individuals $i$, and $\rho = 4$. The superposed line corresponds to $\epsilon_i = 0.5 - a_i$ or the belief for all individuals to be in a median position of the income distribution. Except for the boundary regions close to the minimum and maximum income, the theoretical fit approximates the trend in the data reasonably well. This indicates that the trend to the median is indeed present for the vast majority of the population.}\label{fig:errorplot}
	\end{center}
\end{figure}

An immediate corollary of the population perceiving themselves to earn the median income is the tendency for rather poor individuals to overestimate their position and the rich to underestimate it, as all perceive themselves to be in the middle. Hence, the fit for the median perception tracks the trend in the simulations reasonably well for the vast majority of observations (cf. Figure~\ref{fig:errorplot}). As we discuss in more detail in~\ref{app:percdec}, there is no tendency to the median for the left and right tail of the distributions which the simulation results reflect, too. Indeed, approaching the minimum or maximum improves the accuracy of individual estimates. The intuition for this is quite simple: The poorest and the richest individual will always correctly perceive their social position, independent of $\rho \in \mathbb{R}^{+}_0$. The rationale for this is that the actual minimum (maximum) of the whole will always be the minimum (maximum) of any potential non-empty subset of the population. Apart from such boundary effects, however, we indeed replicate stylised fact $(ii)$ insofar as the poorer half of the population seems to overestimate their social position, while the richer half underestimates it. This finding is in line with the empirical evidence and suggests that total whole population tends to underestimate the degree of inequality, as we will show in the upcoming subsection.

\subsection{Perceived Individual Inequality}
\begin{figure}[!h]
	\begin{center}
		\includegraphics[width=.8\linewidth]{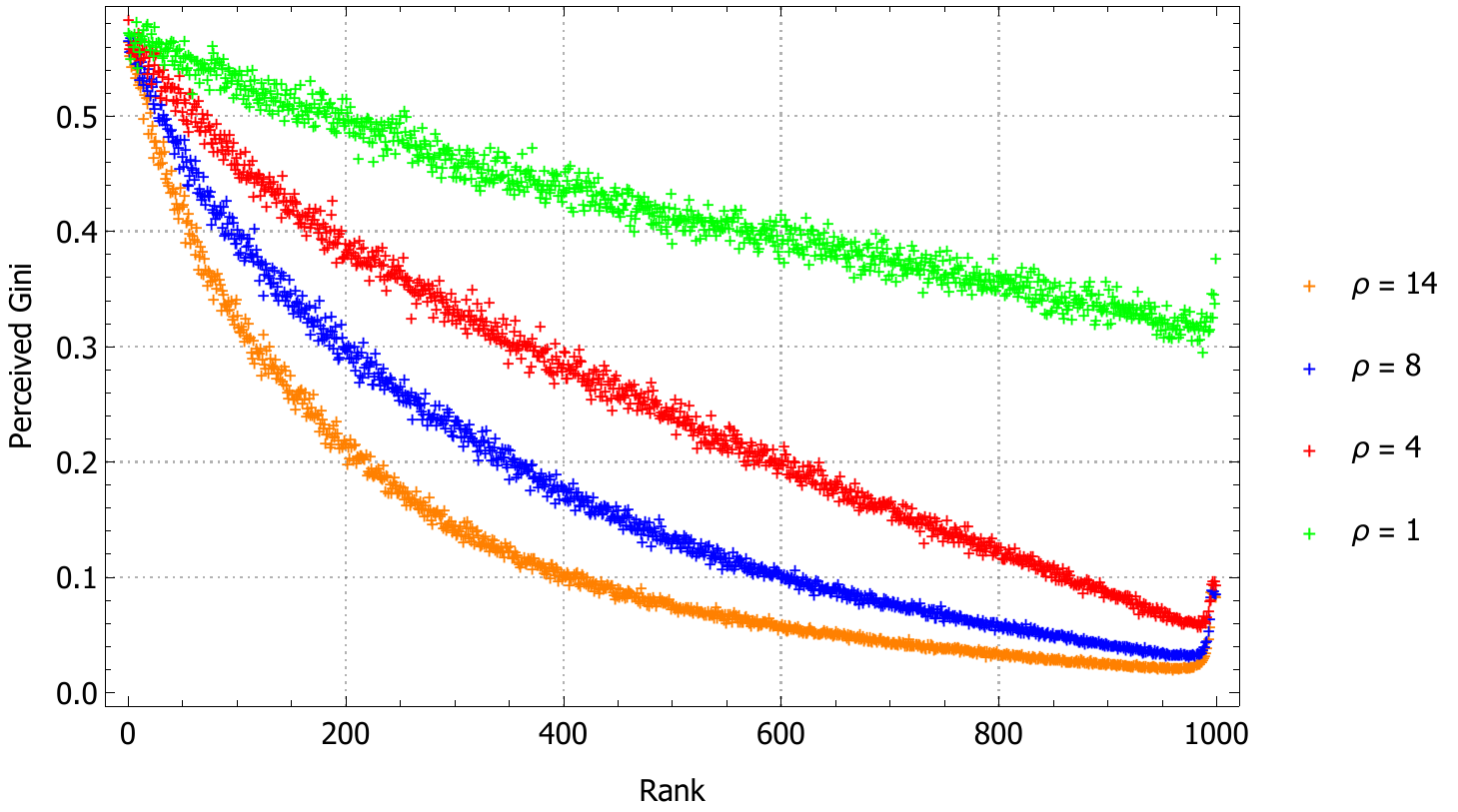}
		\caption{Plot of inequality perceptions against the income rank. Almost all individuals underestimate true inequality with a Gini of $0.5$. Degrees of underestimation vary, though, where bias increases approximately monotonically in income rank. The intuition for this is that homophilic graph formation lets unweighted inequality (absolute income differences) increase only linearly in income rank, but the reference standard (mean income) increases exponentially.}\label{fig:perceptionplot}
	\end{center}
\end{figure}

We define perceived inequality as the Gini coefficient calculated over the perception set of a given individual $i$. In Figure~\ref{fig:perceptionplot}, we plot those perceived Ginis against the income ranks of our individuals with a higher rank indicating a higher income. In line with stylised fact $(iii)$, we find that inequality perceptions decrease approximately monotonically in income rank, while almost all individuals underestimate the actual degree of inequality significantly. As a result of our homophilic graph formation process, perceptions are most accurate for the poorest which either over- or underestimate actual inequality of $G = 0.5$ slightly. The Gini coefficient is conventionally defined as the ratio of (unweighted) mean differences in the incomes within the perception set of an individual to twice the mean income within this group. Homophilic graph formation now lets those unweighted mean differences increase linearly at most, while the mean incomes increase exponentially due to the exponential distribution by which incomes are initialised. As a result, the ratio falls almost monotonically. This results is not only plausible due to its accordance with stylised fact $(iii)$ but might also correspond with the empirical evidence on perception formation. One of the most prominent hypotheses on perception formation from stimuli is the Weber-Fechner law \citep{fechner1862} which indicates that perceived differences in stimuli need to be proportional to the baseline of a given stimulus to be recognisable. The phenomenon is well-established not only for sensory stimuli \citep{formankiewicz2009, pienkowski2009} but also finds use in marketing research on price responses \citep{sirvanci1993, snell1995}. In this framework, one can also understand a decreasing perceived Gini as the change in stimuli (the unweighted differences in incomes of the perception set) do not increase in the same way as the baseline of stimuli (the mean incomes of this perception set) and is thus also consistent with the psychological microevidence.

\subsection{Perceived Global Inequality}
\begin{figure}[H]
	\begin{center}
		\includegraphics[width=.8\linewidth]{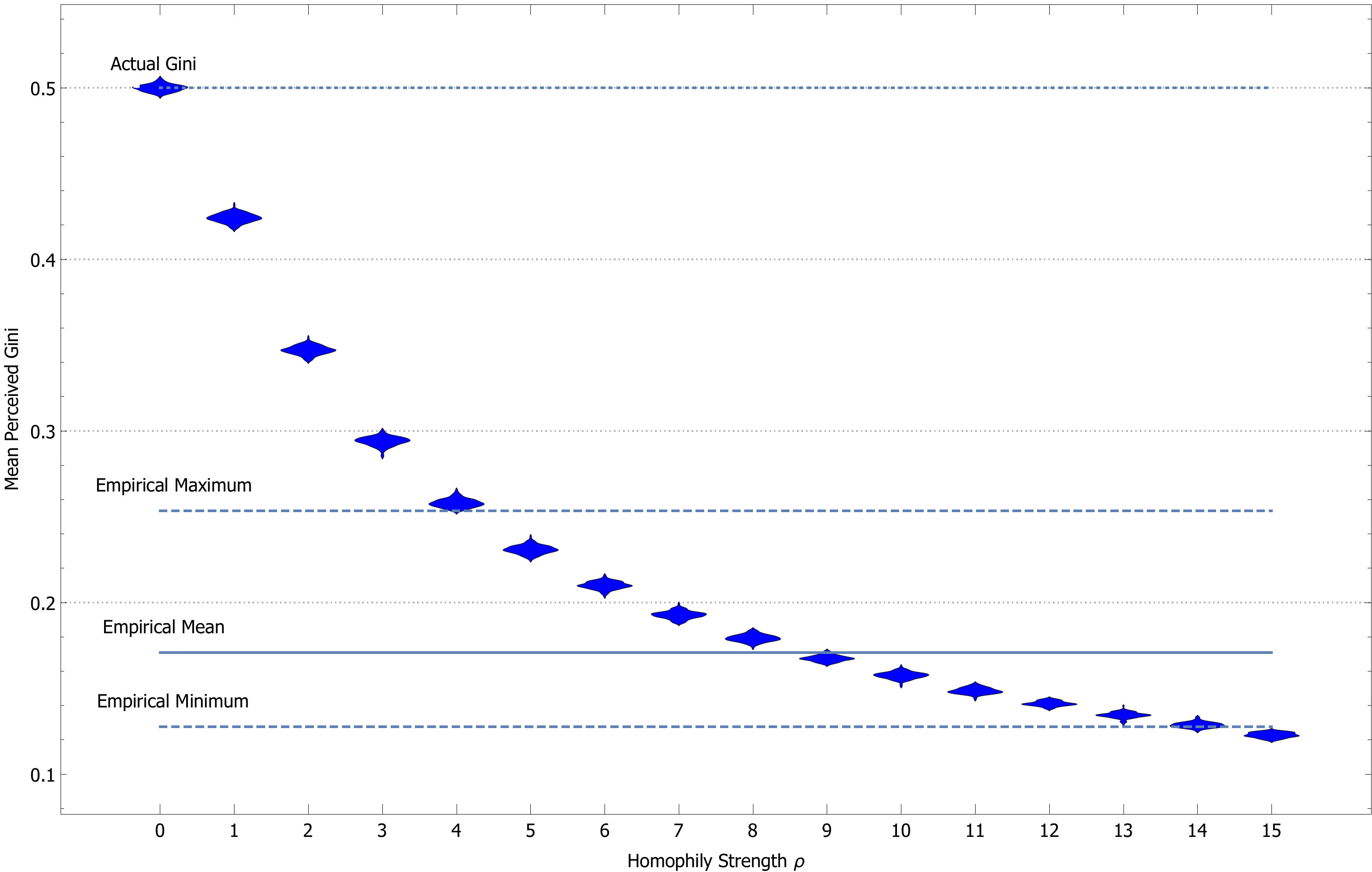}
		\caption{The figure shows the violin plots for the cross-sectional average of individual inequality perceptions per Monte Carlo run of our graph model. The actual unbiased Gini of $G = 0.5$ is indicated by the dotted line. The dashed vertical lines correspond to the empirical sample minimum and maximum, while the bold line corresponds to the sample mean. We find that varying the homophily $\rho$ parameter can fully quantitatively account for the variation in empirical perceptions. }\label{fig:global}
	\end{center}
\end{figure}
For further validation, we also examine whether our graph generating process can \emph{quantitatively} replicate empirical perception patters. We use the mean, minimum and maximum for inequality perceptions on a national level calculated yearly for a large sample of $32$ OECD countries in a $30$ year time-span by \cite{Choi2019}.\footnote{For details and descriptives of their sample, cf. \cite{Choi2019}, especially Appendix B2.} Over all countries, they find a minimum perceived Gini of $G_{min} = 0.1276$, a mean perceived Gini of $G_{mean} = 0.1708$ and a maximum perceived Gini of $G_{max} = 0.2534$. In analogy to their empirical results, we average over the Gini perceptions of all individuals. As we show in Figure~\ref{fig:global}, we find that our process can fully account for their empirical findings and the variation between inequality perceptions by only varying the homophily parameter $\rho$. We also note that the sample average of national inequality perceptions implies a homophily degree $\rho \in [8; 9]$. Yet, our findings imply considerable cross-country variation in homophily that ranges between $\rho \approx 4 $ to $\rho \approx 14$.

\subsection{Perception Dynamics}\label{subsec:perc}
\begin{figure}[!h]
	\begin{center}
		\includegraphics[width=1\linewidth]{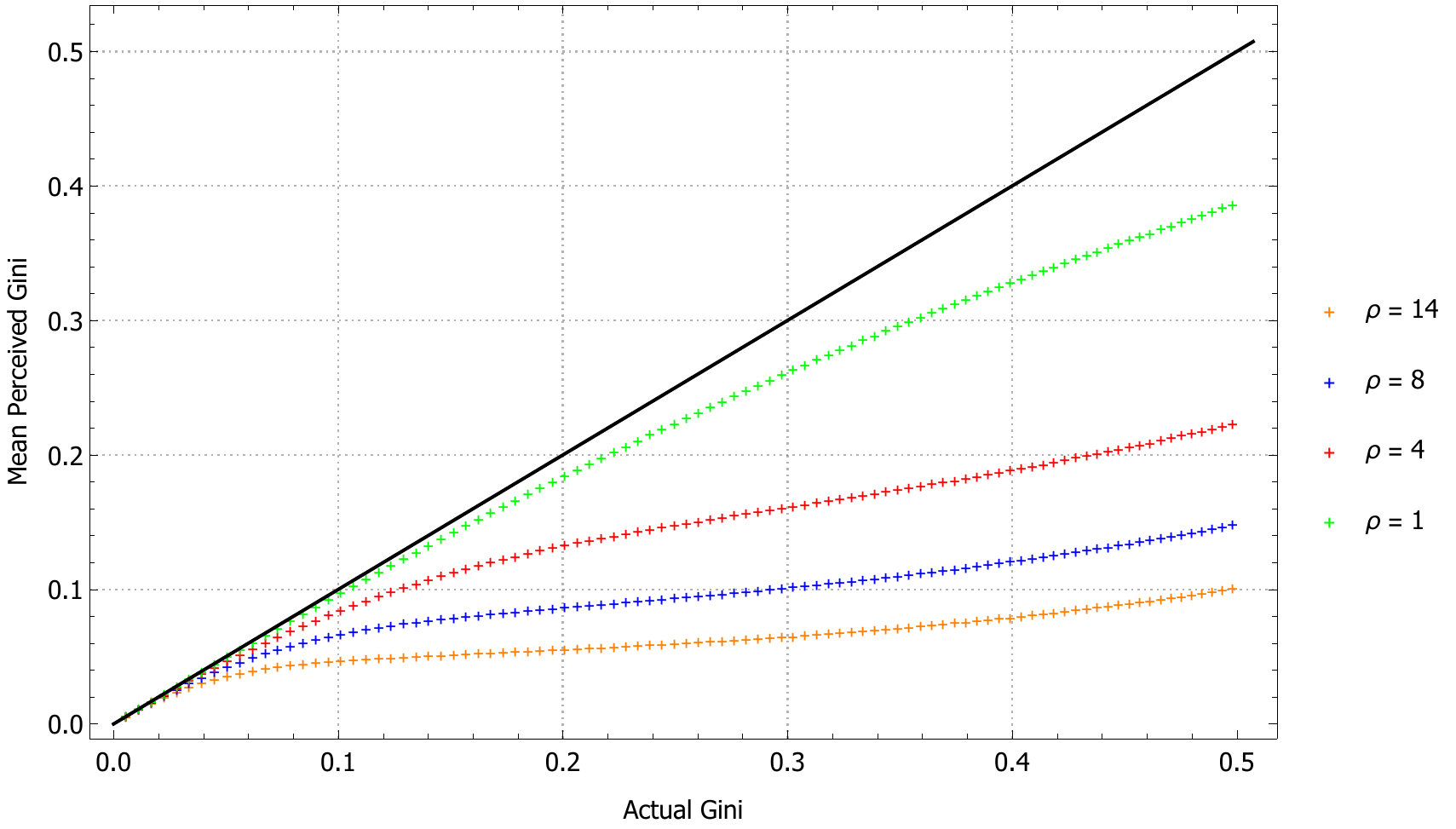}
		\caption{The figure plots varying degrees of true inequality against perceived inequality. The solid $45^{\circ}$ line corresponds to the true Gini coefficient for direct visual comparison. Apart from extremely low homophily level, increases in actual inequality induce a much lower increase in perceived inequality. For moderate and large degrees of homophily, the schedule reaches a plateau rather quickly, i.e., perceived inequality responds extremely slowly to changes in actual inequality. }\label{fig:dynamicplot}
	\end{center}
\end{figure}

To analyse perception dynamics, that is, the reaction of inequality perceptions to changes in actual inequality, we need to initialise the model with another distribution, as the exponential has a fixed Gini of about $0.5$, irrespective of its precise parametrisation. We use the log-normal as another benchmark and vary the dispersion parameter $\sigma$ to simulate changes in the Gini coefficient which is another distribution typically used to describe the skewed nature of empirical income distributions \citep{Knell2020}. As we see, apart from implausibly low degrees of inequality, changes in actual inequality cause far less than a one-to-one change in perceived inequality. Especially for higher degrees of homophily, we find that the schedule quickly reaches a plateau, where inequality perceptions are now extremely persistent with respect to increases in actual inequality. Our model thus is consistent with stylised fact $(iv)$ as our last test of validity.

%There are two main counteracting mechanisms causing this behaviour: First, for any given set of perception sets with fixed identities, an increase in actual inequality also leads to a monotonous increase of inequality in this group and thus, for perceived inequality of the individual in question. Second, however, group compositions are not invariant with respect to inequality. Indeed, our mechanism suggests that 
The persistence in perceptions occurs because homophily becomes more binding and segregation stronger when actual inequality increases. This mechanism leads ceteris paribus to a decrease in perceived inequality which offsets a direct impact of objective inequality on subjective perception.
Compare, for illustration, the two regimes close to a completely egalitarian income distribution near $G = 0$ and relatively high degrees of inequality near $G = 0.5$. The egalitarian state is close to a random network, as homophilic segregation presupposes income differences.
Small changes in actual inequality are thus not strongly reflected in segregation and almost fully impact perceived inequality, leading to a one-to-one correspondence of perceived and actual inequality in this neighbourhood. For large degrees of actual inequality and large homophily, changes in actual inequality immediately impact segregation, leading to a plateau and very persistent perceptions. Notably, this mechanism is not only consistent with the empirical evidence in terms of its emergent outcome; several recent studies by \cite{reardon2011, chen2012} and \cite{toth2021} examine the mechanism directly and show that economic inequality tends to increase (spatial) segregation. A fruitful avenue for further research could be the time-scale on which this channel works, with more laggard segregation responses obviously decreasing the space for inequality-enhancing policies.

\subsection{Segregation Patterns}\label{subsec:seg}
\begin{figure}[!h]
	\begin{center}
		\includegraphics[width=.9\linewidth]{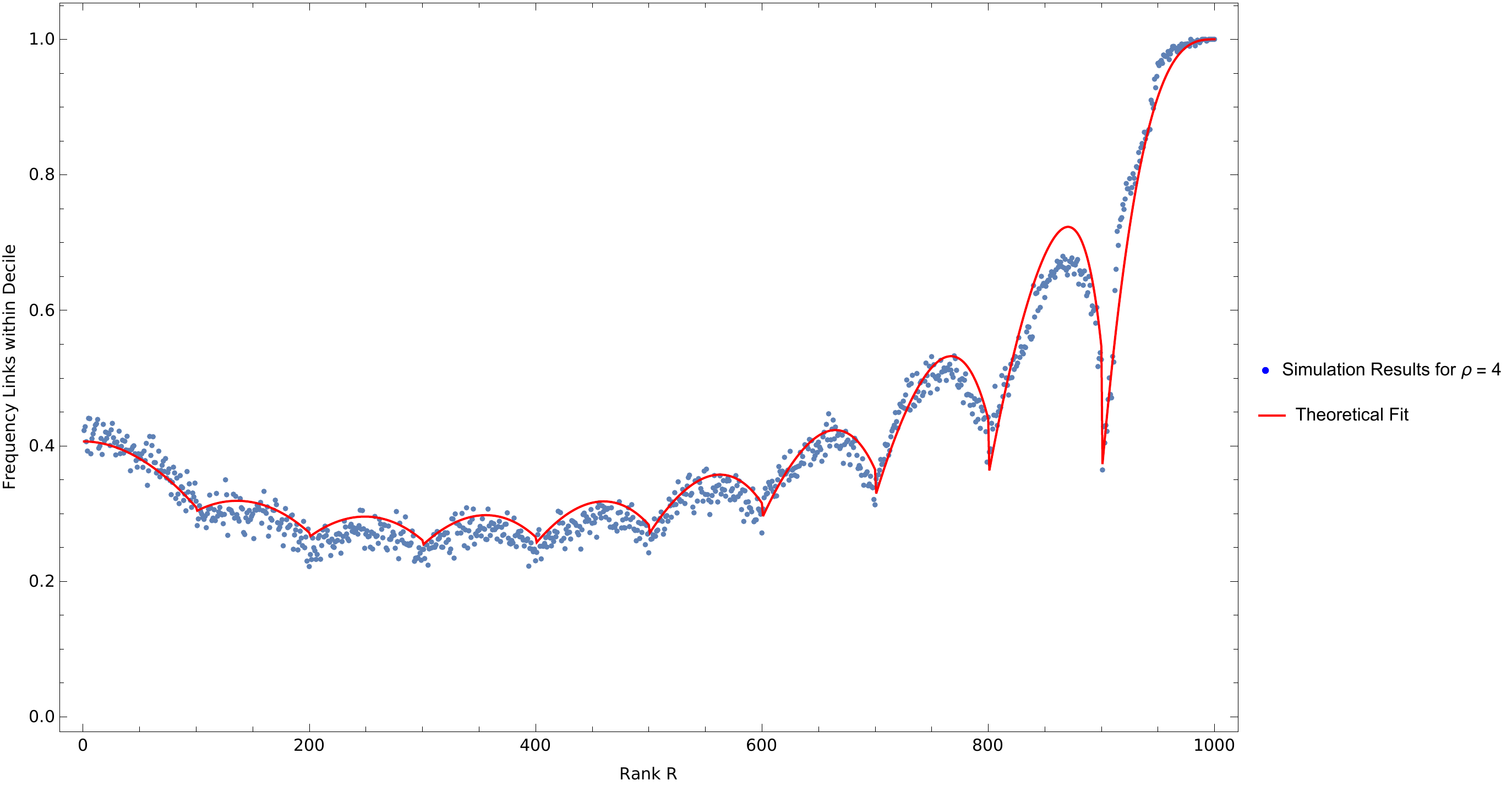}
		\caption{The figure plots our segregation measure, measured as the proportion of links of a node to nodes in the same decile, along the rank distribution. The theoretical fit is obtained for the assumption that nodes choose their neighbours themselves without other nodes choosing links incoming to them. The goodness of fit shows that this is indeed a reasonable assumption. Segregation exhibits distinct and non-trivial patterns both regarding global as well as local maxima.}\label{fig:segregation}
	\end{center}
\end{figure}

We measure segregation as the proportion of links an individual $i$ has in their own decile as $\Delta_i$ as one particular way to measure `selectivity' without access to behavioural linkage parameters. This constitutes a straightforward but standard way to measure segregation and is easily transformed into normalised measures of segregation like the E-I index $\Xi_i$ which is defined as the difference between the share of between-group links and the share of within-group links \citep{Bojanowski2014}.\footnote{Both measures are simple linear transformations of each other due to $\Xi_i = 1 - 2 \Delta_i$.} Figure~\ref{fig:segregation} plots the simulated segregation statistics as well as a theoretical fit for $\rho = 4$. For analytical convenience, the superposed red line plots the probabilities that an individual chooses another agent to link to within their own decile as a first pick, so the total choice set consists of $999$ other individuals, and does not account for the possibility that other agents already link to the agent in question, in contrast to our algorithm. \ref{app:linkage} details the derivation. The goodness of fit demonstrates that these incoming-links do not exhibit a significant effect on segregation patterns and tend to average out in the aggregate, showing that our analytical approximation is indeed reasonable.

Segregation exhibits two distinct patterns along the rank distribution. Firstly, we find that segregation exhibits a skewed U-shape and increases, especially for the richest decile, which is almost completely disconnected from the other groups. In this sense, our graph formation process endogenously creates echo chambers for the richest whose information sets do not cover the poorer population at all. This results from the fact that the richest part of the population is extremely selective in choosing their link-neighbours, as we have shown in Section~\ref{sec:Model}. Secondly, we also find a rather strong variation together with local maxima within deciles. This finding might be, however, spurious and a partial artefact of boundary effects at decile boundaries. 
As~\ref{app:percdec} shows analytically, individuals will choose those sets of link-neighbours with the highest probability that are distributed symmetrically around them in rank. Thus, individuals exactly at the decile boundary will most likely select a set with half of their neighbours across the boundary. Individuals closer to the centre of a decile, on the other hand, will by the same token choose with highest probability link-neighbours within their own decile. Arbitrarily pre-defined group boundaries can thus create within-group variability in commonly used indices like the E-I index that nevertheless exhibits desirable statistical features at an aggregate level \citep{Bojanowski2014}. These findings extend well beyond income deciles, as variables like age group, place of living, gender, education or ethnicity are likely strongly correlated with income. Studies using E-I type indices to detect homophily in other variables might hence create spurious results if income homophily is also present. The relevance of such boundary effects has increasingly also been recognised in applied work \citep{hvidberg2020}. Whenever dimensions have a cardinal scale like income, it might therefore prove more fruitful to use a rolling-window type of estimation, where within-groups are defined in relation to the individual in question, such as a fixed number of income ranks or a fixed income rank interval around theirs. 

\subsection{Summary of Generating Mechanisms}
\begin{figure}[!h]
	\begin{center}
		\includegraphics[width=.8\linewidth]{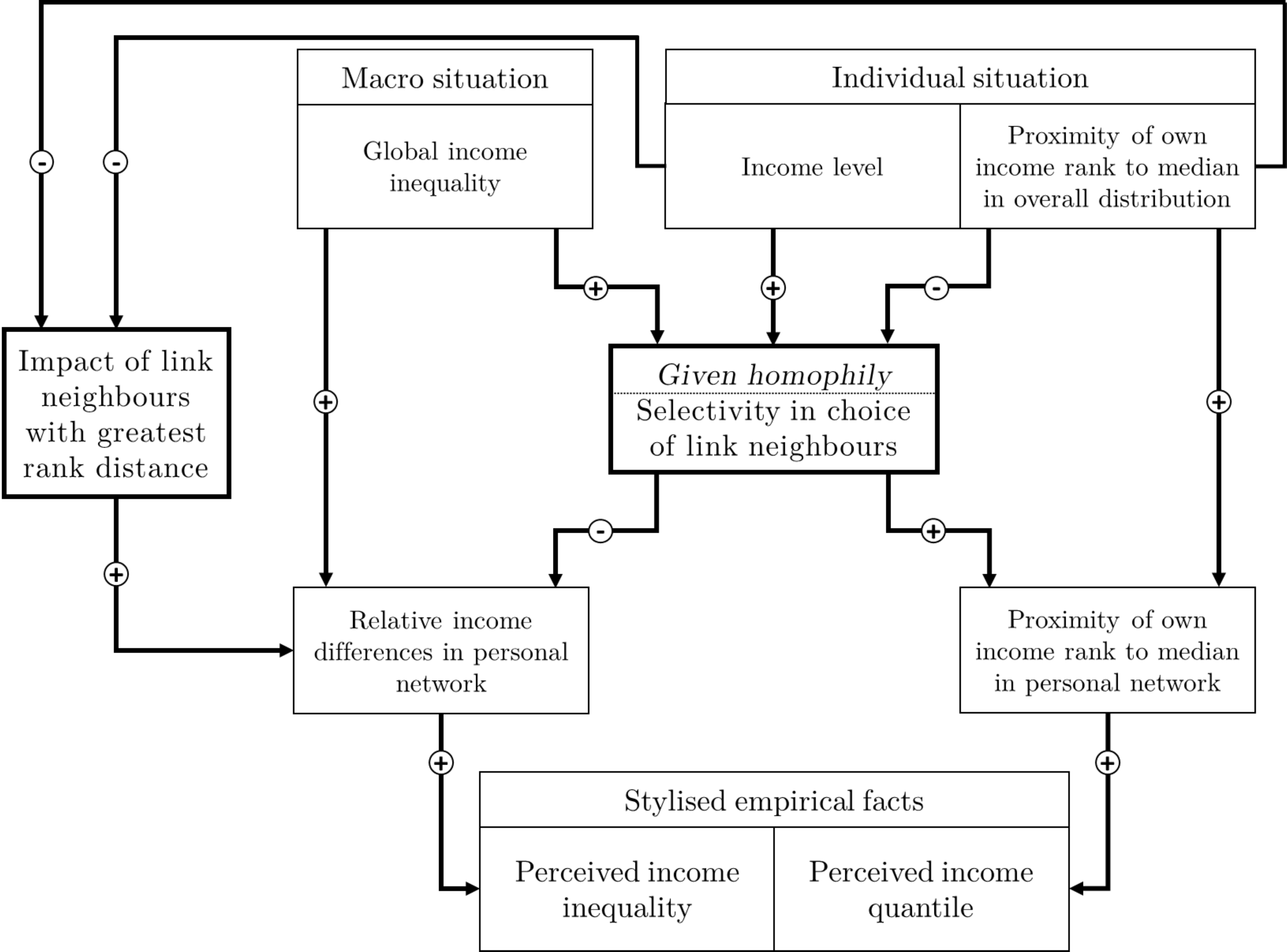}
		\caption{Causal pathway depicting the mechanisms of individual selectivity in choice of link-neighbours and subjective perception in the model given a left-skewed income distribution}\label{fig:causalpathway}
	\end{center}
\end{figure}

For non-biased individual samples, one would expect agents to both estimate global inequality correctly in aggregate and also locate their true income quantile. However, homophily triggers link selectivity and hence biased samples which in turn causes inequality perceptions based on an agent's income level and rank (cf. Figure~\ref{fig:causalpathway})

Selectivity in link formation depends on global income inequality. Furthermore, agents whose income is further from the global median income are more selective in their link-neighbours and so are agents with the higher income ranks, as an artefact of the cut-off in the exponential selection function at the low end of the income distribution (cf. the skewed U-shape in Figures~\ref{fig:pdfrho1} and \ref{fig:pdfrho4}). Such link-neighbour selection generates personal networks for each agent in which this agent tends to have the median income and where income rank differences are relatively small.

Moreover, the extent of relative income differences in one's personal network now depends on the characteristics of the agent in question. Firstly, the impact of link-neighbours with great rank differences to the perceiving agent is larger if this perceiving agent and consequently the majority of link-neighbours have a low income themselves since local inequality calculations weight income differences by the local mean. Secondly, agents close to the global income distribution median ceteris paribus perceive lower inequality levels, for the greatest rank distances tend to be smaller in these cases.

Overall, in a homophilic linking regime, the complex interaction of actual income inequality and of individual absolute income and distribution rank cause biases in income level self-rating and inequality assessment that aggregate to biased underestimation of inequality. Due to the interplay of factors that feed into individual selectivity in choice of link-neighbours, the relation between objective income structure and individual perceptions is non-monotonic and not trivial but requires case-based assessment.

\section{Discussion}\label{sec:Discussion}
Our parsimonious model provides an epistemically possible how-possibly explanation of the stylised empirical facts regarding inequality perceptions that we identified in the literature. Individuals who evaluate their immediate social environment without bias can misperceive their own rank in the overall income distribution as well as global inequality. Homophilic formation of the immediate environment suffices to fully explain the discrepancy between actual and perceived inequality since a rising level of actual inequality causes higher selectivity in link-formation. Moreover, the further away someone's income rank is from the global median and the higher their income, the more selective they are in their choice of link-neighbours. %Hence, the less substantial the differences they encounter when assessing these link-neighbours` incomes
%underestimation of income inequality.

Thus, public misperceptions are not necessarily driven by limitations in information processing, e.g. a behavioural tendency to place oneself near the median of social hierarchy, but by limited information sets the individuals exhibit for inductive reasoning. The seemingly subtle distinction between constraints on information processing and information sets carries important policy implications: When it is the limits of available information and not limits in cognitive ability driving misperceptions, informational treatments may be successful, as also the empirical literature suggests \citep{Cruces2013,mccall2017, Karadja2017}. Such treatments can either consist of delivering information about income inequality itself or facilitating the formation of more diverse contacts in order to overcome the segregation by income that our model finds. In other words, this means breaking up echo chambers that are caused by humans drawing confidence in their beliefs only from repeated observations while ignoring a potential lack of diversity in sources \citep{foster2012,schwarz2016}.

Educating individual citizens about their information deficit and providing ways of overcoming it is important from a democracy theory perspective. For example, \cite[p. 224]{rawls2005} requires ``presently accepted general beliefs'' as basis for arguments in the public forum. However, while one can asses the income inequality objectively without any room for disagreement if using all globally available information, citizens who work only with their individual information will agree on a belief about the Gini that underestimates its actual value. Hence, the lack of individual information access inhibits deliberation about the level inequality and its changes, e.g., in response to past policy measures.

For the first time, we are able to infer the composition of these reference groups from readily available observational data on perceived inequality to inform both empirical investigations as well as more comprehensive model-building in other regards. Directly investigating perception networks might provide a possible remedy for the problem that identification of interaction effects within survey data in the form of time series is hindered by sample sizes that are typically one or two orders of magnitude too low to distinguish noise from true interaction \citep{alfarano2012}. Our main empirical prediction is that the homogeneity of social groups, the fraction of links to agents within the own income decile, roughly follows a U-shaped pattern with a massive decrease in diversity for the richest and (a quantitatively much more modest one) for the poorest agents. This finding might point to an endogenously emerging `elite discourse' with almost no transmission of information to the poorer $90$\% of the population.

Our cross-country analysis shows that there exists considerable variability in implied homophily levels. There are several possible candidates to explain this variability, such as cultural norms, diversity in media and political representation or spatial segregation. Regarding spatial segregation patterns, Thorstein Veblen made the farsighted observation as early as 1899 that urbanisation should increase diversity in social contacts, since cities are the place ``where the human contact of the individual is widest and the mobility of the population is greatest'' \citep[p. 66]{veblen2001}. Thus, perceived inequality should ceteris paribus be higher in urban areas resulting from the higher average income diversity per perception network, a testable hypothesis and thus a possible avenue for further research in spatial economics. This is also what the rather scarce existing evidence for Vietnam and Central and Eastern Europe suggests \citep{mahajan2014, binelli2016}. In this way, spatial and perception network segregation might therefore overlap and interact \citep{Newman2018, toth2021}, and policy affecting the allocation of land could  thus also exhibit unintentional effects on perceptions. We leave detailed analyses on these determinants for further research.

In terms of theory, our empirically validated random geometric graphs might provide an ideal microfoundation for theories of consumption as dependent on the relative income position and for which shocks that affect local income compositions lead to `expenditure cascades' \citep{Duesenberry1949, veblen2001, Frank2014}. This new mechanism might shed light on the disputed link between economic inequality and growth. We will address these questions in further research. 

Finally, our model presents a way of generating random geometric graphs defining both the distribution of the feature that determines linking probability between any pair of nodes and a minimum degree value for each node. Put differently, we apply a Barab{\'a}si-Albert (\citeyear{barabasi1999}) Preferential-Attachment type procedure, which is intuitive for social scientists, to features other than degree and get network graphs that can be analysed using readily available methods from random geometric graph theory. To the best of our knowledge, there is no such way yet.

While it is reassuring that a scale-transformation to a log-scale delivers the same salient `middle-class bias' as our baseline specification, log-transformations in general strike us as a very parsimonious way to capture scale-dependence in choice. This is not only relevant for inequality perceptions but for essentially all variables and features where perception of stimuli is plausibly dependent on scale. In general, it is possible to apply the generating procedure to features other than income that exhibit different distributional patterns. This approach will hopefully inspire future studies of expectation formation, e.g., regarding inflation or business sentiment. In these fields, identification of the relevant perception networks might be a crucial step to bring macroeconomic theory currently mostly building on atomistic rational expectations and empirical studies, that find little support for those types of expectations, closer together \citep{pesaran2006}. Hence, we also provide a toolkit for analysing the impact of homophily regarding any specified feature on network generation (and potential interactions on the resulting network) given a particular distribution of this feature that is well-grounded in the behavioural economics literature. In sum, we anticipate our theoretical framework to generate numerous avenues for further studies both theoretically, regarding graph- and expectation formation, as well as empirically, pertaining to the determinants of homophily and possible policy measures, to information transmission and to the effects of inequality on aggregate consumption. 

\setcounter{section}{0}
\renewcommand\thesection{Appendix \Alph{section}}
\renewcommand\thesubsection{\thesection.\Alph{subsection}}
\section{Linkage Probabilities for Homophilic Networks}\label{app:linkage}
\begin{HD}{} Linkage Probabilities.
	
	Consider an arbitrary node $K_i$ indexed $i \in 0, ..., N-1$ as their rank $R$ increasing in income that is part of a graph $G$ with $N$ nodes characterised by adjacency matrix $A$. Let $I_i$ denote their income, where $f_\lambda (\cdot)$ defines the PDF of a exponential probability distribution defined over the real half-line $(0,\infty)$ with parameter $\lambda > 0$ and $F_\lambda(\cdot)$ the corresponding CDF. The quantile function for any population share $p$ and with parameter $\lambda > 0$ for an exponential distribution is given as
	
	\begin{align}
		\phi_\lambda (p) = \frac{- \log [1 - p]}{\lambda}.
	\end{align}
	We assume without loss of generality that $\lambda = 1$ for normalisation. Calculated $\rho$ values therefore need to be scaled up by the inverse of the mean income, $\lambda^{-1}$ for empirical application.
	
	The quantile of a node with income $I_i$ can be approximated by their rank $R$, such that $p \approx R/N$ as a discrete approximation of the continuous probability density which holds for large $N$. We want to derive the probability that a node $i$ with rank $R \in \mathbb{N}^+_0$ connects to a node $j$ with a distance of $d$ to node $i$.  Expressing the weights as defined in Section~\ref{sec:Model} now in the form of quantiles, we get
	
	\begin{align}
		w_{ij} &= \exp [ -\rho (R/N) - \phi_{\lambda = 1} ((R+d)/N)|]\\
		&= \exp [ -\rho~|\log ( 1- (R/N))  - \log( 1 - (R+d)/ N)|].
	\end{align}
	Assume first that $d> 0$, that is, $j$ is richer than $i$. Simplifying the weights yields for $d > 0$:
	
	\begin{align}
		w_{ij} 	&= \exp [ -\rho~\big(\log ( 1- (R/N)) - \log( 1 - (R+d)/ N))\big)]\\
		&= \exp [\log \big(\frac{N-R}{N}\big)^{-\rho} -\log \big(\frac{N-R-d}{N}\big)^{-\rho}]\\
		&= \big(\frac{N-R}{N-R-d}\big)^{-\rho}\\
		&= \big(\frac{N-R-d}{N-R}\big)^{\rho}
	\end{align}
	
	Analogously, we get for $d < 0$
	\begin{align}
		w_{ij} 	&= \exp [ -\rho~\big(\log ( 1- ((R+d)/N) - \log( 1 - (R/ N))\big)]\\
		&= \exp [\log \big(\frac{N-R-d}{N}\big)^{-\rho}) -\log \big(\frac{N-R}{N}\big)^{-\rho}]\\
		&=  \big(\frac{N-R-d}{N-R}\big)^{-\rho}\\
		&= \big(\frac{N-R}{N-R-d}\big)^{\rho}
	\end{align}
	
	To translate $w_{ij}$ into probabilites, we need to normalise by all weights. Note that this is still a (close) approximation of the probabilities of link-formation of a given node $i$. Nodes draw their $C$ link-neighbours from the set of all neighbours. This implies that draws are not independent, as we assume here. Since $C \ll N$, however, the effect is marginal. The approximation for the probability below, however, seems to perform quite well which we verify in our subsection on segregation. The probability $p$ that $i$ chooses $j$ as a link-partner can therefore be approximated as\\
	
	$p_{ij} (N,R,d) \approx \left\{
	\begin{array}{ll}
		\Big(\big(\frac{N-R}{N-R-d}\big)^{\rho} \big/ (\sum_{\tilde{d}=-R+1}^{-1}~\big(\frac{N-R}{N-R-\tilde{d}}\big)^{\rho} + \sum_{\tilde{d} = 1}^{N-R-1} ~\big(\frac{N-R-\tilde{d}}{N-R}\big)^{\rho}\Big)  & \textrm{for}~d< 0, \\
		\Big(\big(\frac{N-R-d}{N-R}\big)^{\rho}\big/ (\sum_{\tilde{d}=-R+1}^{-1}~\big(\frac{N-R}{N-R-\tilde{d}}\big)^{\rho} + \sum_{\tilde{d} = 1}^{N-R-1} ~\big(\frac{N-R-\tilde{d}}{N-R}\big)^{\rho}\Big) & \, \textrm{for}~d>0.\\
	\end{array}
	\right. $\\
	
	Notice that the function behaves as expected and is monotonically decreasing in $|d| \in \mathbb{N}^+$. The strength of selection also increases monotonically in the homophily parameter $\rho$. For $\rho = 0$, we recover the equiprobable case without any decay. The precise functional form of the decay for $\rho \in \mathbb{R}^{+}$ is, however, far from trivial and changes along the rank distribution. The right tail of the correspondent density is always a power transformation of a linear function, whereas the left tail for any given $R$ is a power transformation of function with hyperbolic decay. In this sense, all nodes are more 'selective' regarding individuals that are poorer than regarding the richer part of the population. To see this, compare the decay for the minimum and the maximum of the distribution for $\rho = 1$ as a special case. For $R = 0$, $p_{ij} \propto 1 - (|d|/N)$ with linear decay in $|d|$, as there exists only a right tail, while for $R = N - 1$, $p_{ij} \propto 1/(1+|d|)$  which decays extremely fast in $|d|$ by a power function, as there exists only a left tail here. In this sense, the richest individual is far more 'selective' in choosing their (poorer) link-neighbours than the poorest individual choosing their (richer) ones.
	
	The theoretical expected segregation index we compare against our simulation results can be straightforwardly computed from those probabilities. Let $\delta_{i}$ be the set of nodes that are in the same group as node $i$ such as an income quantile. The probability to connect with a link-neighbour $\tilde{p}_i$ can then again by approximated as
	
	\begin{align}
		\tilde{p}_i (N,R,d) &\approx \sum_{j \in \delta_{i}} p_{ij}.
	\end{align}
	
\end{HD}
\section{Perceived Quantiles in Perception Networks}\label{app:percdec}
\begin{PS}{}Pure Homophily implies a Tendency to the Median in Perceived Quantiles.\\
	
	Consider an arbitrary node $K_i$ indexed $i \in 0, ..., N-1$ in a graph $G$ characterised by adjacency matrix $A$. Let $I_i$ denote their income, where $f_\lambda (\cdot)$ defines the PDF of a exponential probability distribution defined over the real half-line $(0,\infty)$ with parameter $\lambda > 0$ and $F_\lambda(\cdot)$ the corresponding CDF. Let $M$ be the number of links of node $K_i$ with $M$ even. This leaves us with $N - 1  \choose M$ $= S$ possible permutations of link-neighbours. Assume further for $F_\lambda (I_i)$ that it is between $1/2 \cdot M/N$ and $1- (1/2\cdot M/N))$, such that
	
	\begin{align}
		\frac{1}{2}\frac{M}{N} &<F (I_i) < 1- \frac{1}{2}\frac{M}{N}.\label{boundarycondition}
	\end{align}
	
	Let now $\theta_{ij}$ be an arbitrary realisation of a permitted set of incomes of nodes to which $K_i$ linked, indexed by $j$ out of the set of permitted sets $\Theta_i$ with $\Theta_i = \{\theta_{i1}, ..., \theta_{iS} \}$ and size $S$. Assume further that all incomes in $\theta_{ij}$ are distinct. If link formation is independent of $I_i$ as the sole characteristic differentiating $K_i$ from all other nodes, all sets $\theta_{ij}$ of the same size $M$ are equally likely with probability $1/S$ by extension, since $K_i$ connects to any other node with equal probability. This would be the case for both standard preferential attachment models as well as ER random graphs. 
	
	In our model, the probability $p_{ik}$ that $A_{ik} = 1$ depends negatively on the absolute distance $|I_i - I_k|$, such that $\partial p_{ik} / \partial |I_i - I_k| < 0$. By linearity, the probability $p_{ij}$ of node $i$ to have $\theta_{ij}$ as their chosen set of incomes to which she is linked decreases in the sum of absolute differences, that is, $\partial p_{ij} / \partial \sum_{I_k \in \theta_{ij} } |I_k - I_i| < 0$. It follows, that $p_{ij}$ as a local probability of a set of a given length being chosen by homophilic preferential attachment is maximised for a minimisation of  $\sum_{I_k \in \theta_{ij} } |I_k - I_i|$. Since the benchmark without homophily is equal probability of $1/S$ for all sets of a given size $M$, this condition also maximises the global probability that this set is chosen for a given size $M$. Formally, the minimisation problem chooses a set or sets $\theta_{ij}$ such that
	
	\begin{align}
		\argmin_{\theta_{ij}} & \sum_{I_k\in\theta_{ij}} |I_k - I_i|.\label{minproblem}
	\end{align}
	
	It remains to be shown that this minimisation leads to the choice of a set $\theta_{ij}$ for which $I_i$ is the median value. The median requires the same number of values above or below $I_i$ in $\theta_{ij}$. With $M$ links for node $K_i$ of income rank $R$ and $M$ even, this requires $M/2$ values above and below $I_i$. For $I_i$ as the median being minimising for the absolute distances, this requires i) that there exists no node with rank $R + M/2 + 1$ such that their income distance to $K_i$ is less than the income distance from node $K_i$ to the node ranked $R - M/2$. If i) is violated, the node with rank  $R + M/2 + 1$ is part of the distance-minimising set and thus, $I_i$ is not the median of $\theta_{ij}$. The symmetrical condition ii) requires that there is no node with rank $R - M/2 - 1$ such that its distance to $K_i$ is less than the distance of $K_i$ to the node with rank $R + M/2$. In terms of a quantile function, we require
	
	\begin{align}
		\phi \Large(\frac{R + M/2 + 1 }{N}\Large) - \phi \Large(\frac{R}{N}\Large) &>  \phi \Large(\frac{R}{N}\Large) -  \phi \Large(\frac{R - M/2}{N}\Large)
		\intertext{and}
		\phi \Large(\frac{R}{N}\Large) -\phi \Large(\frac{R  - M/2 - 1 }{N}\Large)   &>   \phi \Large(\frac{R + M/2}{N}\Large) - \phi \Large(\frac{R}{N}\Large).
		\intertext{Rearranging yields}
		\phi \Large(\frac{R + M/2 + 1 }{N}\Large) +  \phi \Large(\frac{R - M/2}{N}\Large) &> 2 \phi \Large(\frac{R}{N}\Large) > \phi \Large(\frac{R + M/2}{N}\Large) + \phi \Large(\frac{R  - M/2 - 1 }{N}\Large).\label{totalinequalities}
	\end{align}

	Expressing the left-hand side of inequalities for a generic distribution in~\eqref{totalinequalities} for an continuous exponential such that $R/N \approx p$ and substituting the quantile function, we require
	
	\begin{align}
		\frac{- \log [1 - (R + M/2 + 1)/N]}{\lambda} +\frac{- \log [1 - (R - M/2)/N]}{\lambda} > 2 \frac{- \log [1 - (R/N)]}{\lambda} \label{firstinequality}
	\end{align}
	
	The condition $R/N \approx p$ presupposes $N$ to be sufficiently large for the discrete realisations of the sample to approximate the quantiles of the continuous exponential distribution. We find this condition fulfilled for several numerical experiments. It is easy to see that the left-hand side condition in~\eqref{totalinequalities} is fulfilled for a quantile function whose first derivative is monotonically increasing which is the case for $d \phi_\lambda (p) /d p = 1/((1-p)\lambda)$ for $p \in [0,1)$ and $\lambda >0$. We can also show this by manipulation of~\eqref{firstinequality} as
	
	\begin{align}
		\log [(1 - (R + M/2 + 1)/N)\cdot (1 - (R - M/2)/N)] &<  \log [(1 - (R/N))^2]\\
		(1 - (R + M/2 + 1)/N)\cdot (1 - (R - M/2)/N) &< (1 - (R/N))^2
	\end{align}
	\begin{align}
		\intertext{which implies}
		1 - \frac{ (R + M/2 + 1)}{N} - \frac{R - M/2}{N} + \frac{(R +  M/2 + 1)(R - M/2)}{N^2} - 1+  2 \frac{R}{N} - \frac{R^2}{N^2} &<0\\
		- \frac{1}{N} + \frac{R - M/2 -(M/2)^2}{N^2} &< 0\\
		\frac{R - M/2 -(M/2)^2 - N}{N^2} &< 0.\label{cond1simplified}
	\end{align}
	Since $R \leq N$ per definition, condition~\eqref{cond1simplified} is trivially fulfilled. Notice that this implies for an exponential initial distribution, $I_i$ cannot be below the median in the most likely set. The right hand-side of inequalities~\eqref{totalinequalities} is a bit more demanding. Stating the condition in terms of the quantile function for an exponential, we get

	\begin{align}
		2 \frac{- \log [1 - (R/N)]}{\lambda} &> \frac{- \log [1 - (R - M/2 - 1)/N]}{\lambda} +\frac{- \log [1 - (R + M/2)/N]}{\lambda}. \label{secondinequality}
		\intertext{Simplifying yields}
		(1- (R/N))^2 &< (1 - (R - M/2 - 1)/N) \cdot (1 - (R + M/2 - 1)/N)\\
		(1- (R/N))^2 &< \frac{(N - R + M/2 + 1)(N - R - M/2)}{N^2}\\
		(N - R)^2 &< N^2 - RN - N M/2 - RN + R^2 + RM/2 + N M/2 - R M/2\nonumber\\
		& - M^2/4 + N - R - M/2\\
		0 &< N - R - M/2 - M^2/4\\
		\frac{R}{N} &< 1 - \frac{1}{2} \frac{M}{N} - \frac{M^2}{4N}.\label{cond2simplified}
	\end{align}
	
	For our discrete sample, $F_\lambda(I_i) \approx R/N$ which reveals that condition~\eqref{cond2simplified} is only a slightly more demanding condition than boundary condition~\eqref{boundarycondition} that guarantees the possibility of $I_i$ being a median in the first place and only differs by $M^2/4N$. Since we typically assume $M \ll N$, this term vanishes. Indeed, for a realistic baseline scenario with $N = 1,000$ and $M = 5$, the condition is fulfilled for the poorest $99 \%$ of the population and thus for the vast majority. Together with the lower boundary condition~\eqref{boundarycondition}, the tendency to place themselves in the middle should exist for about $98$\% of the population and thus the vast majority. Minimising absolute deviations for an exponential income distribution and  $M \ll N$ thus entails choosing sets that let $I_i$ be the median of $\theta_{ij} \cup I_i$ for almost all $I_i$. While the strength of this mechanism will of course be dependent on $\partial p_{ij} / \partial \sum_{I_k \in \theta_{ij} } |I_k - I_i|$, the median is the most likely outcome for any homophilic network as the perceived quantile for the vast majority of nodes.
\end{PS}

\section{Linkage Based on Logarithmised Income Differences}\label{app:logincome}
In this appendix, we discuss the case where agents seek to minimise relative rather than absolute income differences in tie formation, i.e., weights for tie-formation are inversely proportional to the absolute distance in log income. Agent $j$'s weight in agent $i$'s draw $\tilde{w}_{ij}$ is thus determined by

\begin{align}
	\tilde{w}_{ij}&=\frac{1}{\exp[\rho~|\ln [I_{j}] - \ln [I_{i}]|]}.\label{networkformation_log}
\end{align}

The scale-transformation by logarithmising tends to offset the property of the exponential income distribution to exhibit much higher (absolute) income differences in its upper tail than in the lower parts of the distribution, since the natural log has negative second derivative. For moderate to high levels of $\rho$ agents will have a strong tendency to choose an equal number of agents above or below them in income that is roughly homogeneous across the income distribution (apart from agents located at the boundary). For self-perceptions, this implies a `middle-class bias' per stylised facts (i) and, by extension, (ii), as is also readily verified by simulation with the results below in Figures~\ref{fig:denshslog1} to~\ref{fig:denshslog14}.

\begin{figure}[!h]
	\minipage{0.24\textwidth}
	\includegraphics[width=\linewidth]{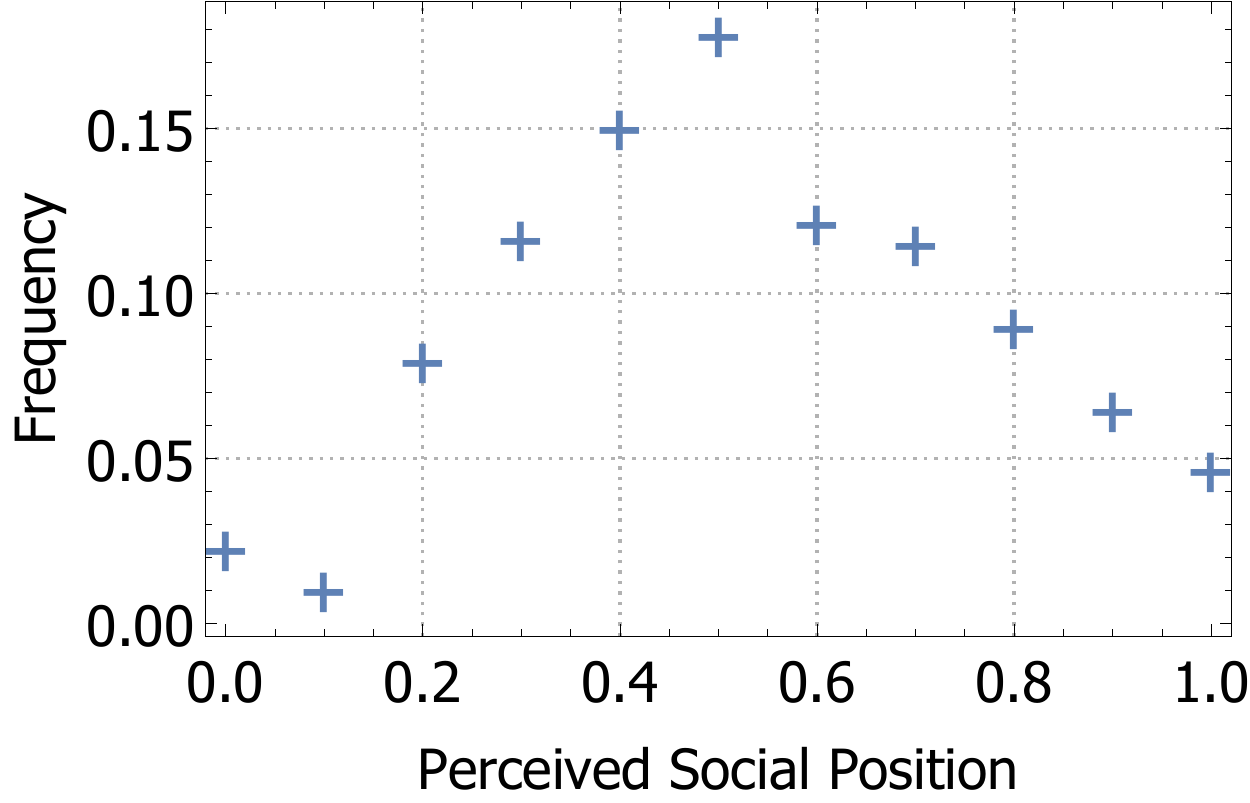}
	\caption{ Density of Perceived Quantiles for $\rho = 1$.}\label{fig:denshslog1}
	\endminipage\hfill
	\minipage{0.24\textwidth}
	\includegraphics[width=\linewidth]{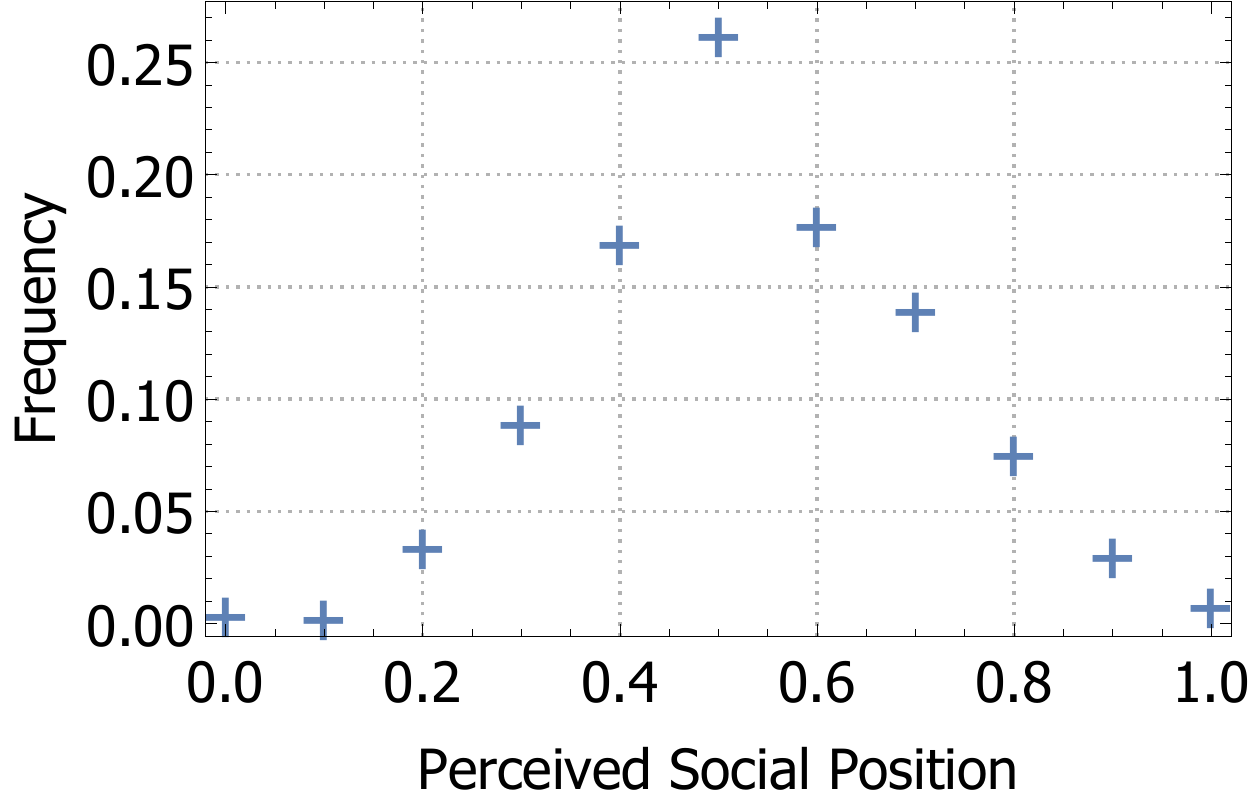}
	\caption{Density of Perceived Quantiles for $\rho = 4$.}\label{fig:denshslog4}
	\endminipage\hfill
	\minipage{0.24\textwidth}%
	\includegraphics[width=\linewidth]{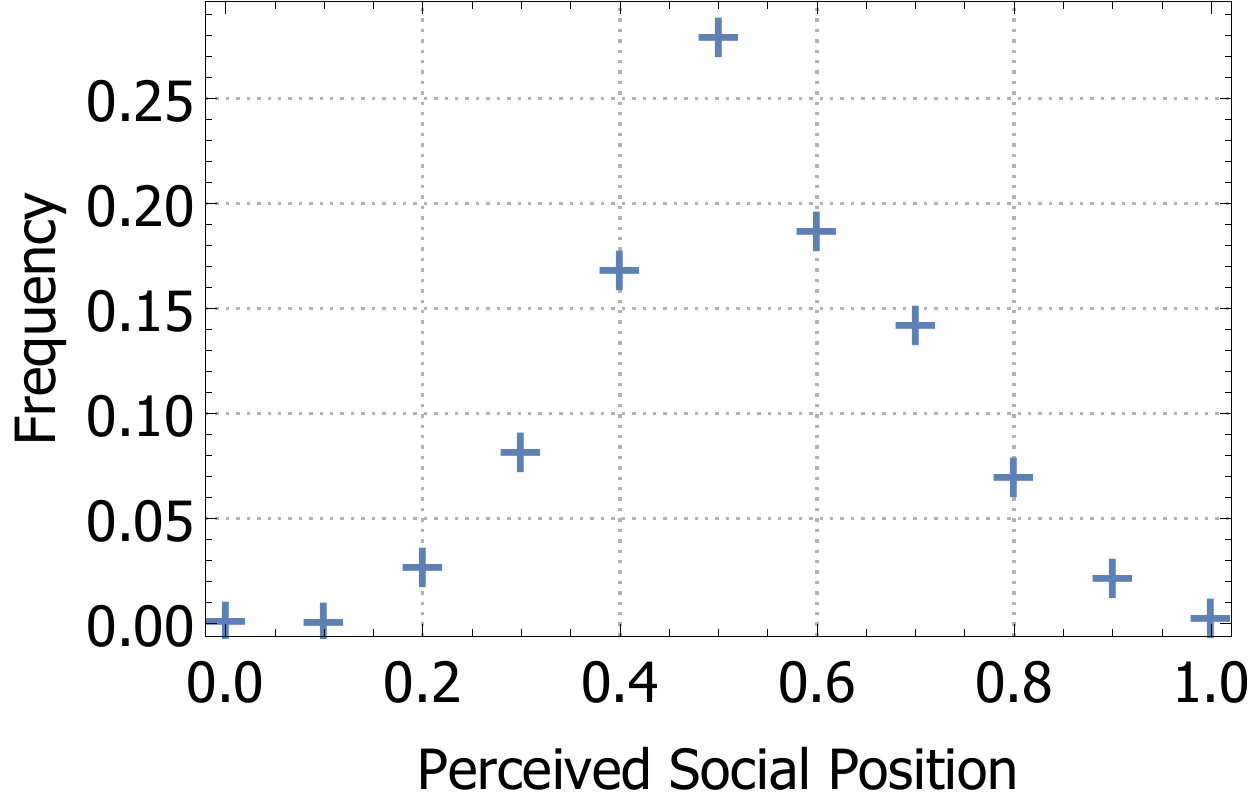}
	\caption{Density of Perceived Quantiles for $\rho = 8$.}\label{fig:denshslog8}
	\endminipage\hfill
	\minipage{0.24\textwidth}%
	\includegraphics[width=\linewidth]{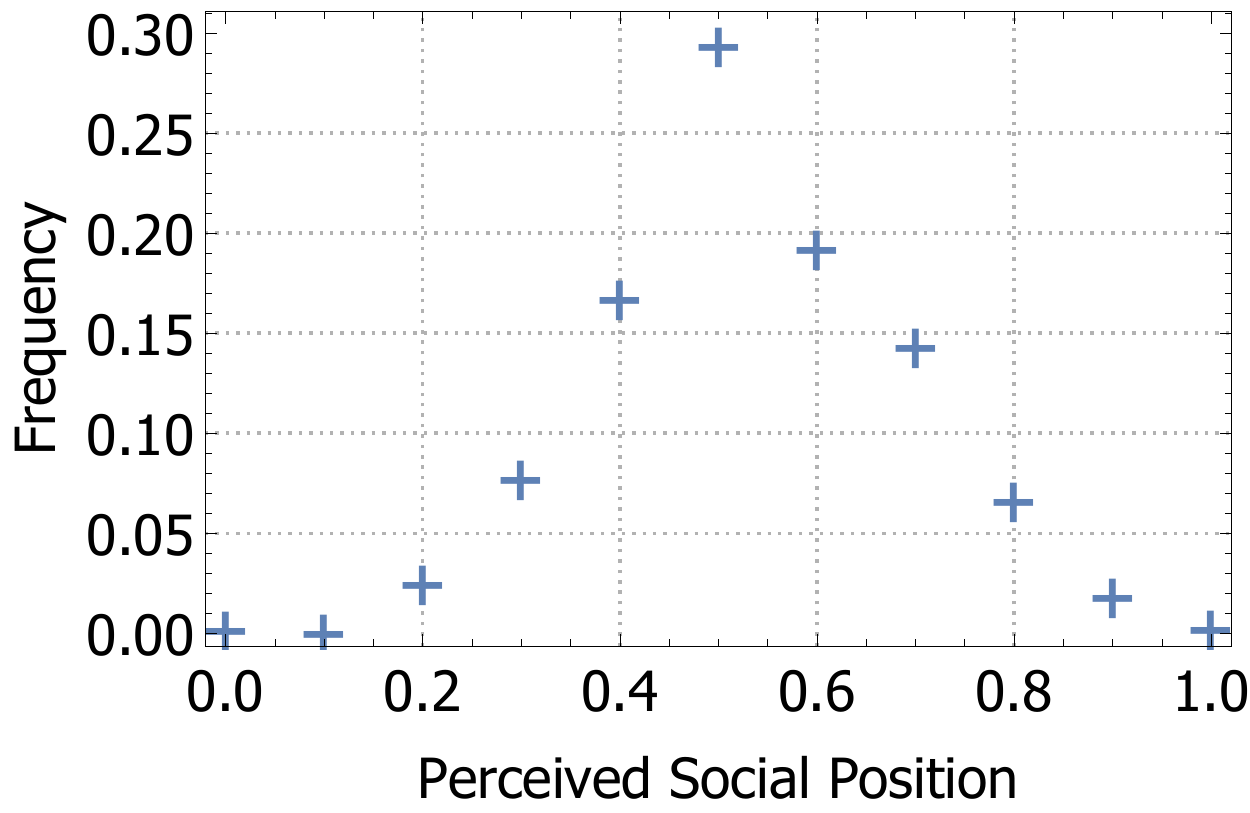}
	\caption{Density of Perceived Quantiles for $\rho = 14$.}\label{fig:denshslog14}
	\endminipage
	\vspace{0.5cm}
	
	\footnotesize{\emph{Note:} The Figures report the perceived social positions for $\rho \in \{1;4;8;14\}$ with $10$ bins each and for a choice function with log incomes in its argument. All Figures exhibit significant deviation from the benchmark with equal frequencies. While the tendency is apparent in all panels, its strength expectedly increases and perceptions are more narrowly clustered around the middle categories the higher $\rho$ is.}
\end{figure}

Logarithmising income in the weight function as in eq.~\eqref{networkformation_log} bears enormous consequences for inequality rather than self-perceptions, though. This is what Figure~\ref{fig:perceptionplot_log} below shows. Apart from agents at the respective upper and lower boundaries, the approximately homogeneous segregation tendency across the income distribution manifests itself in roughly homogeneous inequality perceptions for all considered $\rho \geq 4$. For low $\rho$, the behaviour of local Ginis is thus broadly consistent with one aspect of stylised fact (iii), namely that perceived inequality tends to decrease in income rank, as is also shown in Figure~\ref{fig:perceptionplot_log} for $\rho = 1$. However, poor agents then drastically overestimate inequality in violation of the second aspect of stylised fact (iii) that (almost) all agents somewhat drastically underestimate inequality. 

\begin{figure}[H]
	\begin{center}
		\includegraphics[width=.8\linewidth]{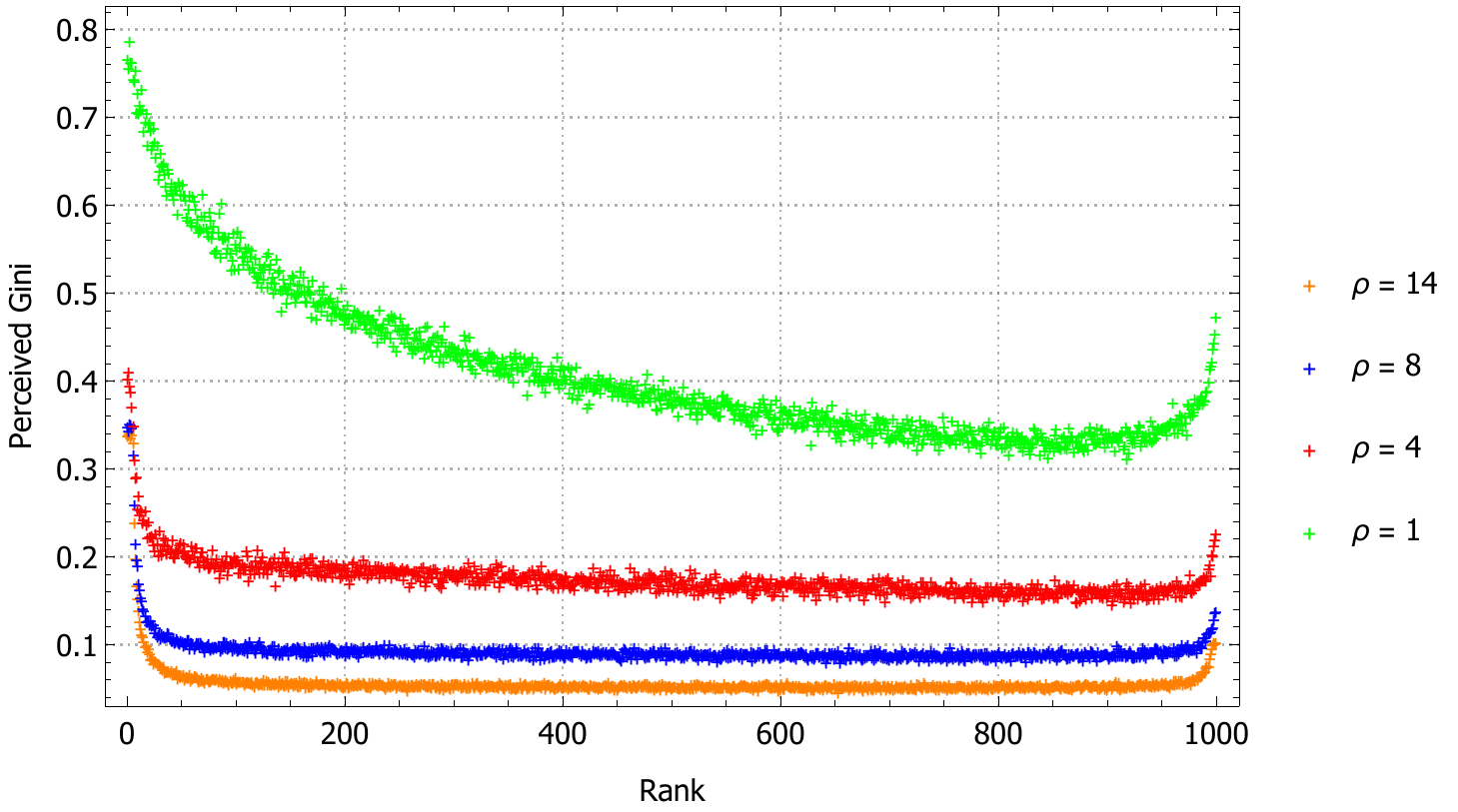}
		\caption{Plot of inequality perceptions against the income rank. Almost all individuals underestimate true inequality with a Gini of $0.5$ in all cases of $\rho > 0$. For moderate to high homophily strenghts $\rho$ between $4$ and  $14$, there is little covariation of income rank and perceptions (except for the inflated perceptions at the upper and lower boundary).}\label{fig:perceptionplot_log}
	\end{center}
\end{figure}

\begin{figure}[H]
	\begin{center}
		\includegraphics[width=.6\linewidth]{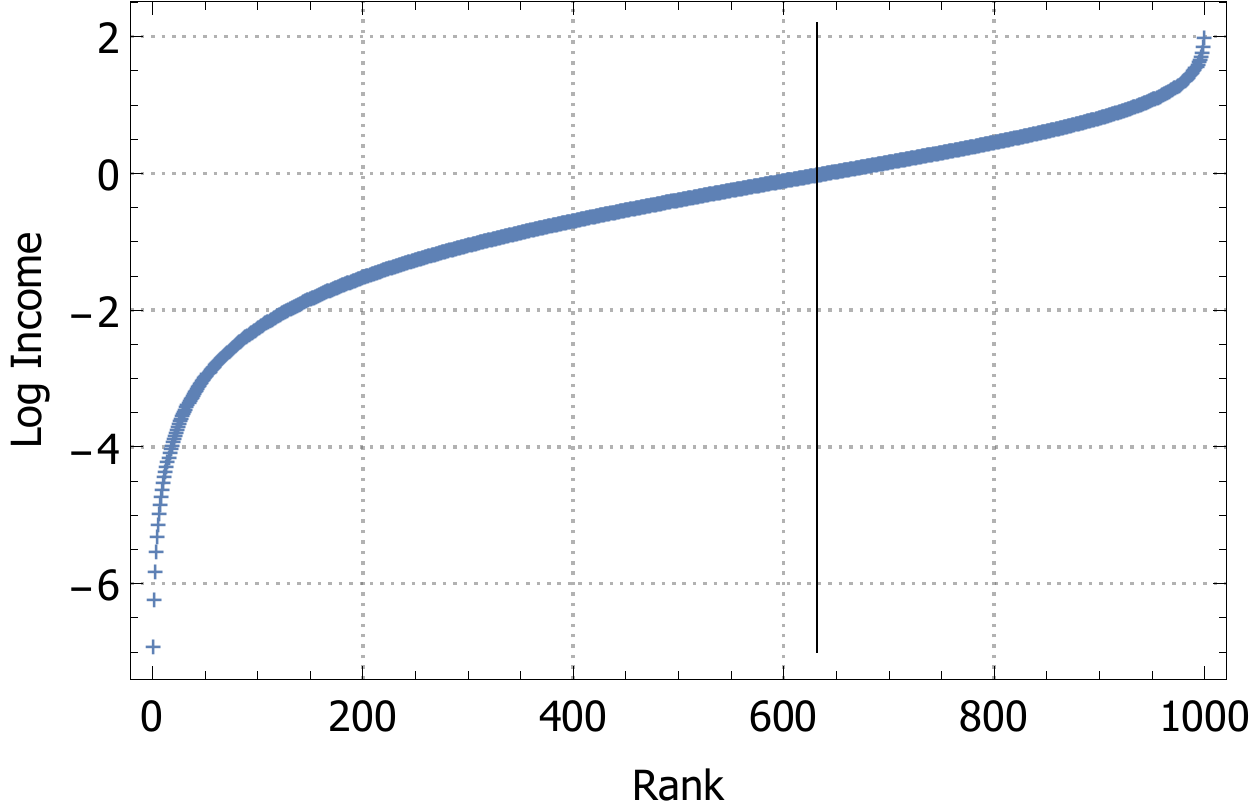}
		\caption{Exponential income distribution in logs as a function of income rank. Around the mean income of $\lambda = 1$, indicated by the vertical gridline, income grows approximately linearly in rank, while growth is superlinear at the tails. }\label{fig:logincomeplot}
	\end{center}
\end{figure}

Heuristically, the behaviour of the local Gini coefficients is a direct consequence of the behaviour of an exponential income distribution on a log-scale. Incomes increase locally linear in rank around the mean income  (of $\lambda = 1$) but increase much faster superlinearly near the upper and lower tails of the distribution, as Figure~\ref{fig:logincomeplot} shows.

This shape of the distribution of logarithmised incomes implies that the agents with income at the lower tail will have a relatively strong probability of observing incomes around the mean, where income ranks do not change inclusion probabilities very much. Since these values are, from the perspective of low-income agents, extreme values, this higher probability of observing incomes within this locally linear region translates into higher expected inequality perceptions, as is evident from the boundary effects in Figure~\ref{fig:perceptionplot_log}. A symmetric argument applies to the upper tail of the distribution, although with a more attenuated effect, since the mean income is closer to the maximum value in income rank due to the skewness of the exponential. Hence, the values around the mean are not as extreme from the perspective of the agents with maximal incomes, leading to a less pronounced increase in upper boundary inequality perceptions. Agents with log incomes located in the locally linear region exhibit the strongest tendency to perceive incomes close to them in rank and exhibit the most pronounced dislike of extreme perceptions at the upper and lower tail. In this sense, the bias against extreme values increases with decreasing distance to the mean income which indicates that inequality perceptions decrease until the mean income is reached and increase afterwards. Increasing $\rho$ then disproportionately affects extreme distances due to the exponential nature of the weights, mitigating any differences in perceptions caused by them. For $\rho \geq 4$, any differential behaviour is then barely visible in local inequality perceptions within Figure~\ref{fig:perceptionplot_log}, in contrast to both stylised fact (iii) and our results for choice with absolute income differences.

% ...

%%%%%%%%%%%%%%%%%%%%%%%%%%%%%%%%%%%%%%%%%%%%%%

% ENDNOTES. Please uncomment the line below in case of notes.
% \theendnotes

%%%%%%%%%%%%%%%%%%%%%%%%%%%%%%%%%%%%%%%%%%%%%%

% REFERENCES.
% The JASSS bibliographic style file (jasss.bst) is included in the bundle. Please use BibTeX, not BibLaTeX.
% Use natbib commands for references (\citep{}, \citet{}, etc.), not standard LaTeX ones (\cite{}).
% Remember to include the doi and url fields in your bib database. The address field should be included for books.
% Please upload the bib file (not just the bbl one) when submitting.

\bibliographystyle{Jasss}
\bibliography{ginibibliography} % Please set the right name for your bib file

%%%%%%%%%%%%%%%%%%%%%%%%%%%%%%%%%%%%%%%%%%%%%%

\end{document}